\documentclass[12pt]{article}

\usepackage{graphicx}
\usepackage{amssymb}
\usepackage{amsmath}
\usepackage{amsthm}
\usepackage{epsf}
\usepackage[parfill]{parskip}
\usepackage[matrix,arrow]{xy}
\usepackage[usenames]{color}
\input{epsf}

\newcommand{\ba}{\begin{eqnarray*}}
\newcommand{\ea}{\end{eqnarray*}}
\newcommand{\ban}{\begin{eqnarray}}
\newcommand{\ean}{\end{eqnarray}}

\newcommand{\be}{\begin{equation*}}
\newcommand{\ee}{\end{equation*}}
\newcommand{\ben}{\begin{equation}}
\newcommand{\een}{\end{equation}}

\newcommand{\bi}{\begin{itemize}}
\newcommand{\ei}{\end{itemize}}

\newcommand{\mbf}[1]{{\boldsymbol {#1} }}

\newcommand{\rk}{{\rm rk\,}}
\newcommand{\ch}{{\rm ch}}
\newcommand{\td}{{\rm td\,}}
\newcommand{\Div}{{\rm div\,}}
\newcommand{\pic}{{\rm Pic\,}}

\newcommand{\Proj}{{\rm Proj\,}}

\newcommand{\IZ}{\mathbb{Z}}
\newcommand{\IC}{\mathbb{C}}
\newcommand{\IP}{\mathbb{P}}
\newcommand{\IN}{\mathbb{N}}
\newcommand{\IR}{\mathbb{R}}
\newcommand{\IQ}{\mathbb{Q}}

\newcommand{\cI}{{\cal I}}

\newcommand{\cN}{{\cal N}}

\newcommand{\cH}{{\cal H}}

\newcommand{\cE}{{\cal E}}
\newcommand{\cO}{{\cal O}}

\newcommand{\cL}{{\cal L}}
\newcommand{\cT}{{\cal T}}
\newcommand{\cF}{{\cal F}}

\newcommand{\fM}{\mathfrak{M}}
\newcommand{\fm}{\mathfrak{m}}

\newcommand{\Hilb}{\mathrm{Hilb}}
\newcommand{\spec}{\mathrm{Spec\,}}

\newcommand{\bX}{\bar{X}}

\def\ii{{\,{\rm i}\,}}
\def\e{{\,\rm e}\,}

\def \nn{\nonumber}

\makeatletter
\@addtoreset{equation}{section}
\makeatother

\newtheorem{Kronheimer_Donaldson}{Proposition}[section]
\newtheorem{ord_loc}{Theorem}[section]

\newtheorem{1cech}{Lemma}[section]
\newtheorem{length_0_dim}{Lemma}[section]
\newtheorem{Hart}{Corollary}[section]

\begin{document}

\begin{titlepage}
\begin{flushright}
LPTENS 09/38\\
IHES/P/09/53\\
HWM--09--10\\
EMPG--09--18\\
\end{flushright}
\begin{center}
\vskip 2cm {\Huge Crystal melting on toric surfaces
\\ \vskip 0.1cm}
\vskip 1cm {\large Michele Cirafici$^{\,a}$, \ Amir-Kian Kashani-Poor$^{\,b}$, \ Richard J. Szabo$^{\,c}$}
\vskip.6cm 

{\it
$^a$ Centro de An\'{a}lise Matem\'{a}tica, Geometria e Sistemas
Din\^{a}micos\\ Departamento de Matem\'atica\\ Instituto Superior T\'ecnico\\
Av. Rovisco Pais, 1049-001 Lisboa, Portugal\\
 \vskip0.5cm}
{\it $^b$ Laboratoire de Physique Th\'eorique de l'\'Ecole Normale Sup\'erieure, \\
24 rue Lhomond, 75231 Paris, France \\ \vskip0.5cm}
{\it
$^c$ Department of Mathematics\\ Heriot-Watt University\\ Colin Maclaurin
Building, Riccarton, Edinburgh EH14 4AS, U.K.\\
and Maxwell Institute for Mathematical Sciences, Edinburgh, U.K.\\ 
\vskip0.5cm}
\end{center} 
\vskip 2.5cm

\begin{abstract}
We study the relationship between the statistical mechanics of crystal
melting and instanton counting in $\cN=4$ supersymmetric $U(1)$ gauge
theory on toric surfaces. We argue that, in contrast to their
six-dimensional cousins, the two problems are related but not
identical. We develop a vertex formalism for the crystal partition
function, which calculates a generating function for the dimension 0
and 1 subschemes of the toric surface, and describe the modifications
required to obtain the corresponding gauge theory partition function.
\end{abstract}

\end{titlepage}
\newpage

\section{Introduction}

The problem of computing instanton contributions to the partition
functions of four-dimensional supersymmetric gauge theories has a multitude of applications in field theory, string theory, and black
hole physics. The algebraic geometry of the corresponding moduli spaces has spawned much interest in
mathematics. In this paper, we will consider a new related counting problem in the
maximally supersymmetric case. Our approach is motivated by the six-dimensional cousin of
the four-dimensional problem.

The vertex formalism~\cite{AKMV,LLLZ} allows for the computation of the
topological string partition function on arbitrary toric (hence
non-compact) Calabi-Yau threefolds. In ref.~\cite{ORV}, this formalism
was recast into an intuitive counting prescription for plane
partitions (three-dimensional Young diagrams), and shown to have the interpretation of a simple statistical mechanics model of crystal
melting. This reformulation was taken as the starting point for
relating topological string theory on toric Calabi-Yau manifolds to a
six-dimensional maximally supersymmetric $U(1)$ gauge theory in
ref.~\cite{INOV} (for a recent discussion of instanton partition functions in various dimensions, see ref. \cite{Nekrasov_Japan}). In as far as the partition function of the gauge theory is the generating function for Donaldson-Thomas invariants, this relationship was proven in refs.~\cite{MNOP,MOOP}.

One now observes that in all examples which have been computed thus far, the $U(1)$ partition
function of $\cN=4$ Vafa-Witten twisted gauge theory in four
dimensions~\cite{VafaWitten} has as prefactor the Euler function $\hat\eta(q)
= q^{-1/24}\, \eta(q)$, the generating function for ordinary
partitions (Young tableaux), raised to the power of the Euler
characteristic of the underlying four-manifold. This suggests that the $\cN=4$ theory might be the four-dimensional analogue of the six-dimensional gauge theory underlying Donaldson-Thomas invariants, with its partition function computable from a melting crystal prescription.

Instantons of four-dimensional $U(N)$ gauge theories on toric surfaces
(not necessarily Calabi-Yau) have been studied in
refs.~\cite{Nekrasov_lgt,FMP,GSST,BonTan,DijkgraafSulkowski,GasLiu,Kool,BPT}. $U(1)$ instantons arise
as building blocks in these works. In ref.~\cite{Nakajima94}, Nakajima
studied $U(N)$ instantons on ALE spaces and showed (see Theorem 3.2 of that paper) that at the fixed
points of an appropriately lifted toric action, they decompose into a
sum of $U(1)$ instantons (see also refs.~\cite{GasLiu,BPT} for
related results). Based on this result, the authors of ref.~\cite{FMP} employed a localization calculation on the explicit ADHM instanton moduli space
to argue that in the case of $\cN=4$ Vafa-Witten twisted gauge
theories on ALE spaces, the $U(N)$ partition function simply
factorizes into $N$ powers of the $U(1)$ partition function -- the rigorous argument for the factorization of the combinatorial problem was provided in ref.~\cite{FujiMin}. They also
indicate a heuristic argument as to why this factorization should hold
in general,\footnote{This is in contrast to the six-dimensional case
  where factorization does not generically hold. See ref.~\cite{CSS} for
an explicit analysis of the Coulomb phase of the six-dimensional $U(N)$ gauge theory
and its relationship to $U(1)$ instantons.} which is supported by calculations in a two-dimensional reduction of the four-dimensional gauge theory on Hirzebruch-Jung spaces~\cite{GSST,BonTan}. In this paper, the $U(1)$ case will be the focus of attention.

In the following, we shall say that an enumerative problem has a melting
crystal description if it can be recast in terms of a box counting
prescription, analogous to that of ref.~\cite{ORV}. We shall see that
in passing from the study of the Hilbert scheme of points, which
features prominently in instanton calculations, to the Hilbert scheme
of curves, we obtain a problem which is a close kin to the gauge
theory problem and has a melting crystal description. We develop a
vertex prescription for calculating the corresponding partition
function, and describe the modifications necessary to arrive at the gauge theory partition function. In the process, we provide
additional motivation for the conjectured forms of the partition functions of $\cN=4$ theory on Hirzebruch-Jung surfaces proposed in refs.~\cite{FMP,GSST}.

The organisation of this paper is as follows. In
Section~\ref{Counting}, we define the enumerative problems which we will address, together with the underlying physical motivation, and introduce the corresponding generating functions. We proceed to compute
the weights that enter in the definition of these generating function in Section~\ref{Weights}. In Section~\ref{vertex}, we set up the
vertex formalism to compute the crystal melting partition function, and work out the explicit examples of the complex projective plane and Hirzebruch-Jung surfaces. Finally, we describe how these partition functions must be modified to arrive at the instanton partition
function of gauge theory in Section~\ref{gauge}. Four
appendices at the end of the paper provide calculational details and background material. In
Appendix~\ref{app_cech}, we compute the Euler characteristics of torus
invariant subschemes of a toric surface directly in \v Cech cohomology. We collect the facts we will need about toric surfaces in general and Hirzebruch-Jung surfaces in particular in Appendix ~\ref{HJ_surfaces}. Appendix~\ref{B} contains a brief review of characteristic classes of coherent sheaves. In Appendix~\ref{cp2}, we illustrate the factorization of the Hilbert scheme of curves into a divisorial and a punctual part, a result which plays a central role in the computations of this paper, for the
example of the projective plane.

We have made an effort to include many intermediate steps and explanatory notes throughout, in the hope of rendering the exposition more accessible to the casual reader.

Unless otherwise noted, all schemes are defined over the field $\IC$.

\section{The enumeration problems\label{Counting}}

In six dimensions, the counting of closed 0 and 1 dimensional
subschemes of a projective scheme $X$ is closely related to a gauge
theory problem on $X$. The crystal melting prescription of
ref.~\cite{ORV}, in hindsight, is most intuitive in this setting, as
the boxes out of which the crystal is built correspond to sections of
the structure sheaf $\cO_Y$ of the corresponding closed subscheme $Y$.\footnote{This hindsight is based on the proof of the equivalence of the generating functions for Donaldson-Thomas and Gromov-Witten invariants on toric threefolds~\cite{MNOP,MOOP}.} In this section, we will explain the parametrization of subschemes and the gauge theory problem in turn. We then discuss why they coincide in six dimensions, and how they are related in four dimensions.

\subsection{The Hilbert scheme } \label{hcofc}

Grothendieck proved that the Hilbert functor of a projective scheme $X$ is representable by a projective scheme called the Hilbert scheme $\Hilb^X_{P(t)}$ of $X$. The closed points of $\Hilb^X_{P(t)}$ correspond to closed subschemes $Y$ of $X$ with Hilbert polynomial $P^X_Y(t) = P(t)$. Recall that upon fixing a very ample line bundle $\cL$, i.e. an embedding of $X$ into projective space, the Hilbert polynomial $P^X_Y(t)\in\IQ[t]$ is defined by the function
\be
P^X_Y(t) = \chi \big(\cO_Y \otimes \cL^{\otimes t}\big) = \sum_{i\geq0}\, (-1)^i\,\dim_\IC H^i(X,\cO_Y\otimes\cL^{\otimes t})
\ee
for sufficiently large $t$.\footnote{This definition makes sense because the right-hand side of this equation is a polynomial for sufficiently large $t$, see e.g.~\cite[Theorem~7.5]{hartshorne}.}
The constant term of the Hilbert polynomial is hence the Euler characteristic of the subscheme $Y$. When $X=\IC\IP^r$ is projective space, the leading term is given by
\be
P^{\IC\IP^r}_Y(t) = \frac{d}{n!}\, t^{n} + \ldots  \,,
\ee
with $n$ the dimension of $Y$ and $d$ its degree.

One can further stratify the Hilbert scheme, e.g. by considering the
Hilbert scheme of curves with fixed homology class $\beta \in H_2(X,\IZ)$. In the case
$X=\IC\IP^2$ reviewed in Appendix~\ref{cp2}, the homology class is
determined by the degree of the curve, and can hence be read off from
the Hilbert polynomial. For space curves, i.e. $X=\IC\IP^3$, this
stratification has been studied in
ref.~\cite{Martin-DeschampsPerrin}. For generic smooth projective surfaces, it is also used in ref.~\cite{DKO}. 

In general, Hilbert schemes are very complicated objects. The Hilbert scheme of points however, for which the Hilbert polynomial is a constant $n$, is well understood. It has already made various appearances in the physics literature (see e.g. \cite{DMV}). It is commonly denoted $X^{[n]}:=\Hilb^X_n$. The Hilbert-Chow morphism 
\be
X^{[n]} ~\longrightarrow~  S^nX  \,,
\ee
with $S^nX$ the $n$-th symmetric product of the scheme $X$, reflects the intuition that away from the locus at which points approach each other, the moduli space of $n$ points on $X$ is simply given by $n$ copies of $X$ modulo permutations.

The Hilbert scheme of curves generally exhibits much richer structure.  On a smooth projective surface $X$, this structure simplifies: codimension 1 subschemes factorize into divisors, i.e. the multiples of integral codimension 1 subschemes, with multiplicity given by their degree, and sums of free and embedded points (see ref.~\cite[p.~514]{Fogarty}, and also ref.~\cite[Section~3]{Stoppa}).\footnote{We thank Richard Thomas for explanations concerning this point.} We will denote the Hilbert scheme of subschemes $Y$ of $X$ with $\beta=[Y] \in H_2(X,\IZ)$ and $n= \chi(\cO_Y)$ as $I_n(X,\beta)$. Given a $Y\in I_n(X,\beta)$, $\beta \in H_2(X,\IZ)$ depends solely on the divisorial part $D$ of $Y$. The contribution of $D$ to $n$ is given by $n_\beta=-\frac12\,\beta\cdot(\beta+K_X)$, with $K_X$ the canonical class of $X$ (see Section \ref{euler_subscheme}). $n-n_\beta$ is due to the free and embedded points $Y_0$ of $Y$. Thus,
\ben
I_{n}(X,\beta)\cong I_{n_\beta}(X,\beta)\times X^{[n-n_\beta]} \ .
\label{Hilbertfact}
\een
To gain some intuition, we illustrate the factorization (\ref{Hilbertfact}) of the Hilbert scheme at the level of the underlying topological spaces explicitly for the toric surface $X=\IC \IP^2$ in Appendix~\ref{cp2}. Let us now consider the two factors contributing to eq.~(\ref{Hilbertfact}). The Hilbert scheme of points $X^{[n]}$ on a smooth projective surface $X$ is non-singular and of dimension $2n$~\cite[Theorem~2.4]{Fogarty}. As for the moduli space of divisors $I_{n_\beta}(X,\beta)$, on a smooth projective surface $X$ of any dimension with $H^1(X,\cO_X) =0$, it is a projective space. It follows that $I_{n}(X,\beta)$ is non-singular \cite[Corollary~2.7]{Fogarty}. This will allow us to define generating functions involving integrals over the fundamental classes $[I_n(X,\beta)]$. Contrary to the six-dimensional case, recourse to virtual classes, as introduced on projective surfaces in ref.~\cite{DKO}, will not be required.

\subsection{Instantons, holomorphic bundles, and sheaves}
\label{instantons}

In the physics literature, instantons are finite-action minima
of a quantum theory. In the case of pure Yang-Mills theory in
four dimensions, they are given by (anti)self-dual connections with
appropriate boundary conditions at infinity. Vafa and
Witten~\cite{VafaWitten} demonstrated that $\cN=4$ supersymmetric
Yang-Mills theory can be modified, following Witten's prescription of topological
twisting, such that its partition function computes, in
favorable circumstances, the Euler characteristic of the moduli space
of instantons. This is guaranteed if certain vanishing theorems
for the geometry of the underlying four-manifold $X$ and gauge bundle $E\to X$ are
met~\cite[Section~2.4]{VafaWitten}, e.g. if $X$ is a compact K\"ahler
manifold of positive curvature and the structure group of $E$ is $SU(2)$. The twisted theory can be encountered
in the wild (i.e. it can describe physical systems) in two situations;
when the manifold $X$ on which the gauge theory is defined is hyperk\"ahler, so that the twisted theory coincides with a subsector of the physical theory, or in certain string theory embeddings, in which the twisting is induced by the background.

For the partition function to be well-defined, a smooth, compact
instanton moduli space is required. One path towards this end is a
compactification given by embedding bundles with irreducible
anti-self-dual connection into the space of semistable sheaves. This
is the Gieseker compactification. A semistable sheaf $\cF$ is in
particular torsion-free.\footnote{The inclusion of torsion-free sheaves in the study of gauge theory instantons was initiated in the physics literature in ref.~\cite{LMNS}. For a relation to non-commutative geometry, see ref.~\cite{NekrasovSchwarz}.} Away from a codimension 2 locus, such
sheaves are locally free, i.e. vector bundles. Hence, intuitively, in passing
from bundles to torsion-free sheaves on surfaces, we are adding pointlike
structures. The precise formulation of this statement is eq.~(\ref{torsion_free}) below.

In this paper, we will solely consider the gauge group $U(1)$. This simple case already merits study for two main reasons:
\begin{enumerate}
\item In six dimensions, this is the gauge group that has been related
  to the topological vertex formalism. 
\item In the case of ALE spaces $X$, ref.~\cite{FMP} performs the
  calculation of the partition function $Z_{U(N)}^{ALE}(X)$ based on
  the description of the $U(N)$ instanton moduli space provided by ref.~\cite{KronheimerNakajima}, and demonstrates that one has the factorization relation
\be
Z_{U(N)}^{ALE}(X) = \big(Z_{U(1)}^{ALE}(X)\big)^{N}  \,.
\ee
The factorization follows from a localization argument which reduces the calculation to the fixed points of an appropriately chosen toric action, and the demonstration in ref.~\cite{Nakajima94} that the instanton bundles on ALE spaces factorize at the fixed points.\footnote{In fact, ref.~\cite{FMP} argues heuristically that this relation should hold in general, as the gauge symmetry of the $U(N)$ theory can be broken to the maximal torus $U(1)^N$ by giving vevs to the scalars, and the partition function is independent of these vevs. This argument is heuristic as the independence of the partition function from the scalar vevs needs to be established carefully.}
\end{enumerate}

Rather than thinking about anti-self-dual connections, we can study holomorphic line bundles. The connection is established by the following well-known proposition.\\

\begin{Kronheimer_Donaldson}
(\cite[Proposition 2.2.6]{DonaldsonKronheimer}) If $H^1(X,\IR)=0$ and
$\cL$ is a line bundle over the surface $X$, then for any 2-form $\omega$ representing $c_1(\cL)$ there is a unique gauge equivalence class of connections with curvature $-2\pi \ii \omega$.
\end{Kronheimer_Donaldson}

Hence, in particular, in mapping line bundles to connections, we can
choose the harmonic representative of $c_1(\cL)$ with respect to a
chosen metric on $X$. The space of harmonic 2-forms $\cH(X)$ on $X$ has
a decomposition into subspaces $\cH^\pm(X)$ of self-dual and
anti-self-dual 2-forms, $$\cH(X) = \cH^+(X) \oplus \cH^-(X) \,.$$ If the
intersection matrix on $H^2(X,\IR)$ is well-defined (a condition we
must impose for non-compact $X$) and negative definite, as will be the case for our main example, the Hirzebruch-Jung surfaces, it follows that $\cH^+(X)=0$, and hence every holomorphic line bundle on such a surface admits an anti-self-dual connection. Since exact forms cannot be anti-self-dual, this connection is unique.

Note finally that even in the case of $U(1)$ gauge theory, where the instanton moduli space is a lattice and in no need of regularization, we continue to consider the larger space of torsion-free sheaves, in accord with points 1 and 2 above.

\subsection{Crystal melting vs. gauge theory}

On a toric Calabi-Yau threefold $X$, the problem that Maulik, Nekrasov, Okounkov, and Pandharipande address in ref.~\cite{MNOP} is the counting of subschemes $Y$ of compact support with no component of codimension 1, and with holomorphic Euler characteristic $n$ and second homology class $\beta$,
\be
n=\chi(\cO_Y)  \,, \quad \beta = [Y] \in H_2(X,\IZ)  \,.
\ee
They denote the corresponding moduli space of ideal sheaves by $I_n(X,\beta)$. They then compute the generating function
\be
Z_{DT}(X;q,w) = \sum_{\beta \in H_2(X,\IZ)} ~ \sum_{n \in \IZ} \,\tilde{N}_{n,\beta} ~ q^n\, w^\beta
\ee
for the Donaldson-Thomas invariants
\be
\tilde{N}_{n,\beta} = \int_{[I_n(X,\beta)]^{vir}}\, 1 \ ,
\ee
the lengths of the 0 dimensional virtual fundamental cycles.
On a projective scheme $X$, $I_n(X,\beta)$ is the moduli space
parametrizing isomorphism classes of torsion-free sheaves of rank 1 with trivial determinant, where the singularity sets of the torsion-free sheaves constitute the subschemes being counted.\footnote{The argument is the following (see e.g. ref.~\cite{OSS}). A torsion-free sheaf $\cT$ injects into its double dual. The determinant sheaf of a rank $r$ torsion-free sheaf is defined to be 
\be
\det \cT = \big(\mbox{$\bigwedge^r$}\, \cT\big)^{**} \,.
\ee
This is a line bundle, as the double dual of a sheaf is reflexive (i.e. isomorphic to its double dual), and reflexive sheaves of rank 1 are locally free. Rank 1 torsion-free sheaves with trivial determinant hence possess an injection into the structure sheaf, i.e. they are ideal sheaves. Note that the distinction between trivializable determinant and trivial determinant is important here. The singularity set $S(\cT)$ of a torsion-free sheaf $\cT$ occurs in codimension 2 or higher, hence the corresponding subscheme has no component in codimension 0 or 1. This argument can be summarized in the exact sequence of sheaves
\be
\xymatrix{
0 ~ \ar[r] & ~ \cT~ \ar[r] & ~ \cT^{**}~ 
\ar[d]^\cong \ar[r] & ~ S(\cT)~  \ar[r] & ~ 0 \\ 
& & ~ \cO_X ~  & &
}
\ee
where the vertical isomorphism is a \emph{fixed} trivialization.

Conversely, the ideal sheaf $\cI_Y$ of a proper closed subscheme $Y$ of a noetherian integral scheme $X$ is a coherent sheaf of rank 1. As a subsheaf of the structure sheaf $\cO_X$, it is torsion-free by the integrality assumption on $X$. If $Y$ has no support in codimension 1, then the determinant of $\cI_Y$ is trivial.}

Shifting focus from six-dimensional Calabi-Yau manifolds to toric surfaces, the counting problem that gives rise to a
melting crystal interpretation and vertex formulation is again that of
counting all subschemes of dimension 0 and 1 of compact support. In
four dimensions, this does not quite map to a counting problem of
torsion-free sheaves. Torsion-free sheaves of trivial determinant on a
surface are the ideal sheaves merely of points, by the same argument
as above. We can enlarge this space in two ways. We can add 1
dimensional subschemes of compact support by hand and consider the corresponding Hilbert
scheme of compact curves, or equivalently the moduli space of ideal sheaves $$I_n(X,\beta)$$ introduced above, now for $X$ a surface.
Alternatively, we can drop the trivial determinant condition, thus enlarging the space to involve factors of line bundles $\cL$, given by the double duals of torsion-free sheaves $\cT$, such that
\ben  \label{torsion_free}
\cT=\cL \otimes \cI_Y  \,.  
\een
Again dimension 1
subschemes come into play, by their relation to effective divisors,
however now only up to linear equivalence. Indeed, the decomposition
(\ref{torsion_free}) corresponds to the gauge theory problem outlined
in Section~\ref{instantons}.

In the case of surfaces, we will hence be computing two different generating functions, defined as follows.

\paragraph{Crystal melting.} 
\ben
Z_{cm}(X;q,w) = \sum_{\beta \in H_2(X,\IZ)}~ \sum_{n \in \IZ}\, {N}^{cm}_{n,\beta} ~ q^n \, w^\beta \,, \label{pcc}
\een
where $n= \chi(\cO_Y)$, $\beta = [Y] \in H_2(X,\IZ)$ and
\ben
N^{cm}_{n,\beta} = \int_{I_n(X,\beta)} \, e\big(TI_n(X,\beta)\big) \,.  \label{ncm}
\een
Here and below, $e(E)$ denotes the Euler class of the bundle $E$. As announced above, no recourse to virtual fundamental classes is taken in these definitions.

\paragraph{Gauge theory.}
\ben
Z_{gt}(X;q,w) = \sum_{\beta \in H_2(X,\IZ)} ~ \sum_{n\in\IQ}\, {N}^{gt}_{n,\beta} ~ q^{-n}\, w^\beta \,,  \label{Zgt}
\een
where $n= \ch_2(\cT)$, $\beta = \ch_1(\cT) \in H_2(X,\IZ)$ and
\ben
N^{gt}_{n,\beta} = \int_{\fM_X(\beta,n)}\, e\big(T\fM_X(\beta,n)\big)  \,, \label{ngt}
\een
with $\fM_X(\beta,n)$ the moduli space parametrizing isomorphism
classes of torsion-free sheaves $\cT$ of the given Chern character. Note that for torsion-free sheaves on non-compact spaces, the second Chern characteristic class $n$ can be fractional; we therefore take the summation in (\ref{Zgt}) over $\IQ$, with $N_{n,\beta}^{gt} =0$ away from a fixed common denominator of $n$, depending on $X$.

Explaining the various ingredients in these formulae, and interpreting and evaluating the integrals (\ref{ncm}) and (\ref{ngt}), will occupy the rest of this paper.

\section{The weights for the generating functions\label{Weights}}

Following the six-dimensional discussion of ref. \cite{MNOP}, we organize the crystal melting counting problem (\ref{pcc}) on a toric surface $X$ in terms of the
holomorphic Euler characteristic $\chi(\cO_Y)$ of the subschemes $Y$, which serves as a weight in the generating function,
and their second homology class $[Y]\in H_2(X,\IZ)$. For the gauge theory partition function,
the weight originates in the action, which for anti-self-dual
connections in four dimensions evaluates to the Chern class of the
vector bundle (locally free sheaf) via Chern-Weil theory. When we
compactify the space of gauge connections by including pointlike
instantons, it is natural to retain the Chern class as weight, as in (\ref{Zgt}). In this
section, we will compute these two weights and find that on Calabi-Yau surfaces, they are equal up to sign, at least for torically invariant $Y$ (this qualification arises due to the non-compactness of $X$, see point 3. in Subsection \ref{gauge_weight}).

\subsection{The Euler characteristic of subschemes} \label{euler_subscheme}

Based on the factorization (\ref{Hilbertfact}), we can calculate the Euler characteristic of $Y$ by adding the contributions from the divisorial and punctual parts, $D$ and $Y_0$, of $Y$: the Euler
characteristic of a 0 dimensional scheme enumerates its global sections,
\be
\chi(\cO_{Y_0}) = h^0(Y_0,\cO_{Y_0}) \,,
\ee
while the Euler characteristic of a divisor $D$ on a surface $X$, as is reviewed in Appendix \ref{B}, is given by
\ben  \label{chioy2} 
\chi (\cO_D)  = -\mbox{$\frac{1}{2}$}\, D\cdot (D+K_X)\,, 
\een
with $K_X$ the canonical class of the surface. The right-hand side of eq.~(\ref{chioy2}) clearly only depends on the class of the divisors up to linear equivalence. We will denote this class by square brackets $[-]$ below. Altogether,
\ben \label{chitot}
\chi(\cO_Y) =  -\mbox{$\frac{1}{2}$}\, D\cdot (D+K_X) + h^0(Y_0,\cO_{Y_0}) \,.
\een

The derivation of formula (\ref{chioy2}) in Appendix \ref{B} relies on the application of the Hirzebruch-Riemann-Roch theorem, valid for $X$ projective. When relaxing the compactness condition, the terms on the left- and right-hand side of eq.~(\ref{chioy2}) remain well-defined for $D$ a divisor with compact support. By calculating the Euler characteristic directly in \v Cech cohomology in Appendix~\ref{app_cech}, following ref.~\cite{MNOP}, we will see that at least in the case of torical\-ly invariant subschemes, eq.~(\ref{chioy2}) remains valid for $D$ of compact support also on a non-compact toric surface $X$. 

To explicitly determine the Euler characteristic of a given divisor on a toric surface $X$, we benefit from the property that the Chow ring $A_1(X)$ is generated by the classes of torically invariant divisors $D_i$ (this is in fact true in arbitrary dimensions). We will enumerate the $D_i$ via $i=0, \dots, n+1$, reserving the indices $i=0$ and $i=n+1$ for non-compact toric divisors if these are present, otherwise setting $D_0=D_{n+1}=0$. This notation allows for the simultaneous treatment of compact and non-compact toric surfaces.

Expanding $[D]$ in classes $[D_i]$ generated by compactly supported divisors, $$[D] = \sum_{i=1}^n\, \lambda_i \,[D_i] \,, $$ with $\lambda_i$ non-negative integers, and with the intersection matrix as given in eq.~(\ref{int_mat}) of Appendix~\ref{HJ_surfaces}, the calculation of the Euler characteristic in Appendix~\ref{app_cech} yields
\ben   \label{chi_cech}
\chi(\cO_{D}) = \sum_{i=1}^n\, \Big(a_i\, \frac{\lambda_i\, (\lambda_i-1)}{2} + \lambda_i - \lambda_i\, \lambda_{i+1}\Big) \,,
\een
where the $a_i$ denote the negative self-intersection numbers $a_i = - D_i^2$, and $\lambda_{n+1}=0$ is introduced for notational convenience.

To compare with eq.~(\ref{chioy2}), note that the total Chern class of a non-singular toric variety
$X$ is given by $$c_t(X)= \prod_{i=0}^{n+1}\, (1 + [D_i])\,.$$ It follows that $$K_X= - \sum_{i=0}^{n+1}\, [D_i]\,,$$ 
and hence
\ben
\chi(\cO_D) = -\frac{1}{2} \sum_{i,j=1}^n \lambda_i (\lambda_j-1) \,D_i \cdot D_j +\frac{1}{2} \sum_{i=1}^n \lambda_i\, D_i\cdot (D_0 + D_{n+1})   \,. \label{chioy_eval}
\een
Borrowing the result (\ref{lin_equ_apq}) from Section \ref{gauge}, in which $[D_0]$ and $[D_{n+1}]$ are expressed as linear combinations of $[D_1], \ldots, [D_n]$, we conclude that eq.~(\ref{chioy2}), though derived for compact $X$, reproduces the \v Cech cohomology result (\ref{chi_cech}) for toric divisors on non-compact $X$ as well.

\subsection{The Chern characteristic of sheaves} \label{gauge_weight}

In the gauge theory setup, we wish to weigh all sheaves via the degree of their second Chern character. This is the natural extension of the notion of instanton number beyond locally free sheaves (i.e. vector bundles), see Appendix~\ref{B}. The Chern character satisfies the multiplicative property
\be
\ch(\cE \otimes \cF) = \ch(\cE) \cdot \ch(\cF) \,.
\ee
For the ideal sheaf $\cI_Z$ of a cycle $Z$, an application of the
Grothendieck-Riemann-Roch theorem yields 
\be
\ch(\cI_Z) = 1 - \eta_Z  \,,
\ee
with $\eta_Z$ the class of the cycle (see e.g. ref.~\cite[p.~159]{HarrisMorrison}). Due to the relation $\ch_0(\cL) = \rk(\cL)=1$ for a line bundle $\cL$, we have
\be
\ch_2(\cL \otimes \cI_Z) = \ch_2(\cL) - \eta_Z \,.
\ee
We can therefore consider the two factors contributing to the weight
of a given torsion-free sheaf separately.

As we show in Lemma~\ref{length_0_dim} of Appendix~\ref{B},
$\deg(\eta_Z) = \chi(\cO_Z)$.

\begin{figure}[h]
 \centering
  \includegraphics[width=5cm]{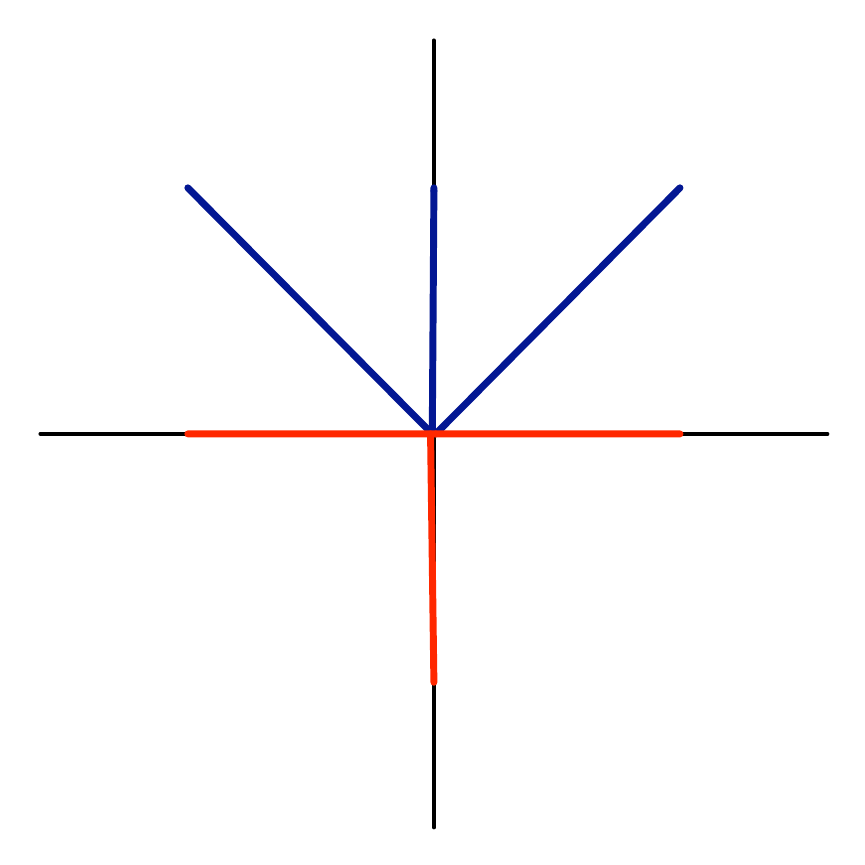}
 \caption{\footnotesize{A toric compactification of the resolved $A_1$ singularity (in blue) to $\IP^1 \times \IP^1$ blown up at two points.}}
 \label{toric_comp}
\end{figure}

For line bundles, we can work in the more familiar cohomological
setup. On a compact surface $X$, the instanton number evaluates to the intersection pairing of the corresponding divisors,
\ben
\frac{1}{2}\, \int_X\, c_1\big(\cO_X(D)\big) \wedge c_1\big(\cO_X(D)
\big) = \frac{1}{2}\, D\cdot D   \,.  \label{instno}
\een
For a divisor $D$ with compact support, this relation in fact continues to hold on arbitrary toric manifolds. The argument consists of three parts:
\begin{enumerate}
\item Also on non-compact manifolds, the first Chern class
  $c_1(\cO_X(D))$ of the line bundle associated to a divisor and the
  closed Poincar\'e dual $\eta_D$ of the support of the divisor are
  cohomologous, $c_1(\cO_X(D)) \sim \eta_D$.\footnote{Recall that the
    closed Poincar\'e dual is integrated against forms of compact
    support, in contrast to the compact Poincar\'e dual which itself
    has compact support (see e.g. ref.~\cite[pp.~51--53]{BottTu}). Hence, the closed Poincar\'e dual $\eta_\Sigma$ of a cycle $\Sigma$ satisfies
\be
\int_\Sigma \, \psi = \int_X \, \eta_\Sigma \wedge \psi
\ee
for any $\psi \in H_c^*(X,\IR)$ (defining both sides to vanish if the form degrees are not appropriate). By restricting the integration to the support of $\psi$, the argument establishing $c_1(\cO_X(D)) \sim \eta_D$ on compact manifolds (see e.g. ref.~\cite[p.~143]{GriffithsHarris}) goes through in the non-compact case.
}
\item Since the support $|D|$ is compact, we can replace the closed Poincar\'e dual by the compact Poincar\'e dual. By localization of this class (in the sense of e.g. ref.~\cite{BottTu}), we know that we can choose its support to be contained in an arbitrary open set containing $|D|$, such that the integral in eq.~(\ref{instno}) is well-defined.
\item We verify the relation (\ref{instno}) by performing the calculation on a toric compactification $\bar{X}$ of $X$ for which the compactification divisor does not intersect the image of the support of $D$. This property guarantees that the evaluations of the integrals over $X$ and $\bar{X}$ coincide. An example of such a compactification, which always exists for smooth toric surfaces, is given in Figure~\ref{toric_comp}.
\end{enumerate}

The toric assumption can be replaced by the requirement that a compactification with the requisite properties exists.

We thus arrive at
\be
\ch_2\big(\cO_X(D) \otimes \cI_Z\big) = \mbox{$\frac{1}{2}$}\, D\cdot D - \chi(\cO_Z) \,.
\ee
For $K_X=0$ this agrees with eq.~(\ref{chitot}) up to sign, as announced at the
beginning of this section.

It turns out that all non-compact divisors in the geometries that we
will consider are linearly equivalent to divisors with compact
support. This thus allows us to compute the instanton number on the
full Picard group. Note that the instanton numbers of line bundles
associated to prime divisors with non-compact support are no longer necessarily integral.

\section{Crystal melting: counting subschemes} \label{vertex}

In the previous sections, we have introduced two closely related enumerative
problems: counting subschemes vs. counting torsion-free sheaves on a
toric surface. The line bundle factor in eq.~(\ref{torsion_free})
requires invoking linear equivalence between toric divisors, and hence
cannot be straightforwardly implemented within a vertex formalism that
essentially only allows for nearest neighbor interactions (in terms of
the 2-cones of the toric fan, or the vertices of the dual web
diagram).\footnote{We say `essentially' as even the vertex formalism
  in six dimensions requires identifying homologous curve classes by
  hand. In fact, when $H^1(X,\cO_X)=H^2(X,\cO_X)=0$, $\pic(X) \cong H^2(X,\IZ)$, so `linearly equivalent' and `homologous' are the same notion on smooth compact toric surfaces (this remains true in any dimension: all cohomology classes on toric manifolds are analytic, hence of pure type $(p,p)$). Even so, in $Z_{cm}$, all curves are counted and only the weight $w^\beta$ invokes homological equivalence, whereas in $Z_{gt}$, the enumeration itself proceeds over equivalence classes.} The problem of counting, in an appropriate sense, the 0 and 1 dimensional compactly supported subschemes of a toric surface does however have a melting crystal implementation, as we demonstrate in this section.

The factorization (\ref{Hilbertfact}) of the Hilbert scheme discussed in Section \ref{hcofc} results in a factorization of the partition function $Z_{cm}$ defined in eq.~(\ref{pcc}),
\ban
Z_{cm}(X;q,w)=\sum_{\beta\in H_2(X,\IZ)}~\int_{I_{n_\beta}(X,\beta)}\,
e\big(T I_{n_\beta}(X,\beta)\big)~q^{n_\beta}\,w^\beta~\sum_{n\geq 0}~\int_{X^{[n]}}\,
e\big(TX^{[n]}\big)~q^n\,. \nn\\  \label{Z_cm_factor}
\ean
We begin by considering the contribution of the free and embedded 0 dimensional subschemes. The corresponding generating function for smooth projective surfaces has been calculated by G\"ottsche in ref.~\cite{Goettsche}. We reproduce his result in the case of toric surfaces via a localization calculation, which then permits an extension to the non-compact case. This calculation will also be relevant in the
gauge theory context of Section~\ref{gauge}. We next apply a similar localization argument to the divisorial contribution to the partition function. Finally, we show how the computation of $Z_{cm}$ can be encapsulated in a small set of diagrammatic rules, and illustrate these in the examples of projective space and Hirzebruch-Jung spaces. Localization arguments lie at the heart of the calculations in this section.

\subsection{0 dimensional subschemes} \label{0dimss}

For $X$ a smooth projective surface, the moduli space $X^{[n]}$ of 0 dimensional subschemes of length $n$ is non-singular of dimension $2n$~\cite[Theorem~2.4]{Fogarty}. The generating function we are after was computed by G\"ottsche for smooth projective surfaces as \cite{Goettsche}
\ben
\sum_{n\geq0}\, \chi\big(X^{[n]} \big)~q^n =  \big(\hat\eta(q)^{-1} \big)^{\chi(X)}   \,, \label{gottsche}
\een
with $$\hat\eta(q)=\prod_{k=1}^\infty\,\big(1-q^k\big)  $$ the generating function of partitions. The computation of ref.~\cite{Goettsche} does not require
a torus action on the surface. If such an action exists, i.e. in the case of a toric
surface $X$, we can reproduce G\"ottsche's formula (\ref{gottsche})
by a localization computation (see also
ref.~\cite{EllingsrudStromme} and ref.~\cite[Appendix~A]{FujiMin}). As the Hilbert scheme $X^{[n]}$ is
smooth, we can use conventional Atiyah-Bott
localization~\cite{AtiyahBott} in equivariant cohomology. We quote
here the more general integration formula of Edidin and
Graham~\cite{EdidinGraham} in equivariant Chow theory; this is the framework that generalizes beyond smooth
varieties.\\

\begin{ord_loc}
\label{ord_loc}
(\cite[Proposition~5]{EdidinGraham}) Let $M$ be a smooth and complete
scheme with the action of a torus $T=(\IC^*)^k$. Denote the fixed point locus of the $T$-action by $M^T$, with embedding $$i: M^T ~ \hookrightarrow~ M \,.$$ Let $a \in A_0(M)$ descend from an equivariant class $\alpha \in A_0^T(M)$, i.e. $a = i^* \alpha$. Then
\ben \label{localization}
\deg(a) = \sum_{F \subset M^T}\, {\pi_F}_* \Big(\, \frac{i_F^* \alpha}{e_T(N_F M)} \,\Big) \,,
\een
where the sum runs through the connected components of the fixed point
locus, $N_FM$ denotes the normal bundle over $F$ in $M$, $i_F$ the
embedding of $F$ into $M$, and $\pi_F$ the projection of $F$ to a
point.
\end{ord_loc}
\bigskip
$e_T(N_FM)$ in formula~(\ref{localization}) denotes the $T$-equivariant Euler class, which is indeed invertible in $A_*^T(F)\otimes_{\IQ[t_1,\dots,t_k]}\IQ[t_1,\dots,t_k]_\fm$ , where
$\IQ[t_1,\dots,t_k]_\fm$ is the localization of the ring
$\IQ[t_1,\dots,t_k]$ at the maximal ideal $\fm$ spanned by the
generators $t_1,\dots,t_k$ of the equivariant ring of~$T$.

When the set of $T$-fixed points is a union of isolated points, the
tangent bundle to each $F$ is trivial, and thus
\be
e_T(N_FM) = e_T(TM|_F) \,.
\ee
With $a=e(TM)$, we have $\alpha=e_T(TM)$, and hence
$i_F^*\alpha=e_T(TM|_F)$ cancels the denominator in the
integrand on the right-hand side of the localization formula (\ref{localization}).
In this case, the integral (\ref{localization}) can simply be evaluated by counting the fixed points of the $T$-action on $M$. 

Note that characteristic classes of equivariant vector
bundles can be extended to equivariant classes. Theorem~\ref{ord_loc} hence
applies to our case of interest, with $M=X^{[n]}$ and $a=e(TX^{[n]})$ the Euler class
of the tangent bundle of the Hilbert scheme $X^{[n]}$.

\begin{figure}[h]
 \centering
  \includegraphics[width=6cm]{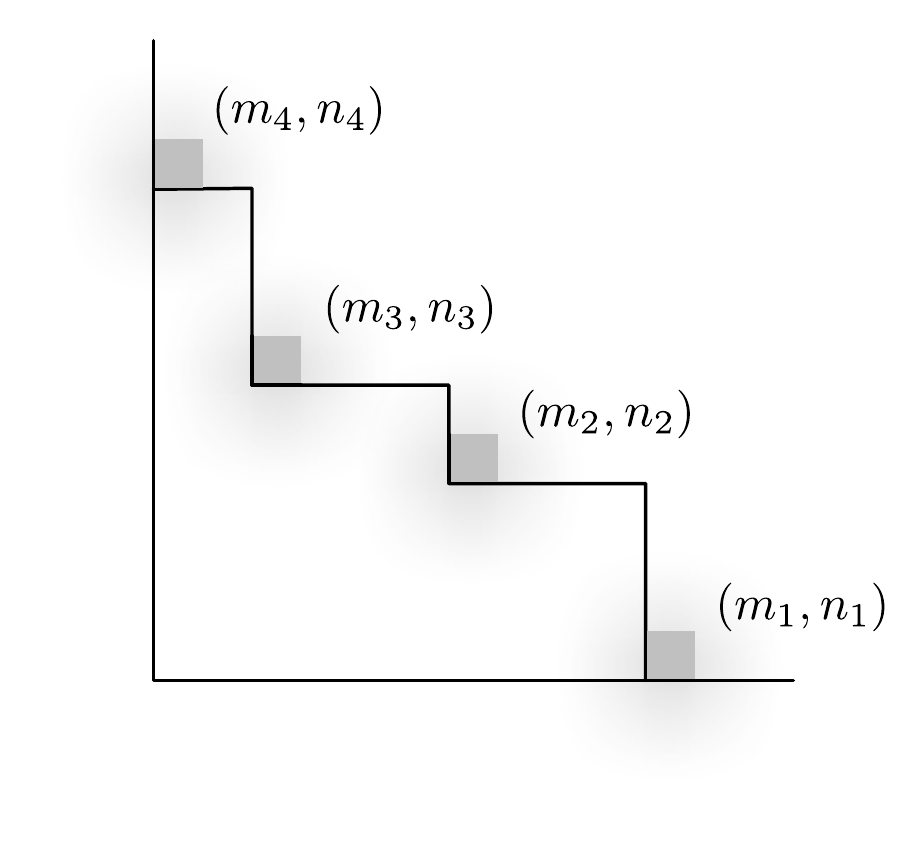}
 \caption{\footnotesize{The Young tableau encoding the ideal generated by the monomials $x^{m_i}\, y^{n_i}$, $i=1, \ldots, 4$, with $(m_i,n_i)$ the coordinates of the shaded boxes.}}
 \label{ideal_partition}
\end{figure}

The fixed points of the torus action on $X^{[n]}$ parametrize
the torically invariant 0 dimensional subschemes of $X$ with holomorphic
Euler characteristic $n$. They can be enumerated by considering an
affine patch $\IC[x, y]$ around each set theoretic fixed point, and
ideals $I\subset\IC[x, y]$ generated by monomials $x^m\, y^n$ giving rise to
non-reduced schemes with support at this point. The ideals $I$ are in one-to-one correspondence with Young tableaux $\pi_I$, as illustrated in Figure~\ref{ideal_partition}.
The boxes of the Young tableau map to a basis of global sections of the corresponding 0 dimensional subscheme $Y$. Its Euler characteristic is hence equal to  
\be
\chi(\cO_Y) = h^0(Y,\cO_Y) = \dim_\IC\big(\IC[x,y]\,\big/\, I\big) =  |\pi_I|
\ ,
\ee
the number of boxes in the Young tableau $\pi_I$. The contribution to the partition function per geometric fixed point is hence
\be
\sum_{\pi} \, q^{|\pi|} = \hat\eta(q)^{-1} \,.
\ee
The toric fixed points correspond to the maximal cones of the toric
fan of $X$. Since the Euler characteristic $\chi(X)$ of a toric
manifold $X$ is given by the number of maximal cones of $X$, this
reproduces the formula (\ref{gottsche}).

The application of standard theorems is complicated when the surface
$X$ is non-compact. We will proceed by applying the localization
formula to a toric compactification $\bX$ of $X$, and then restrict to the fixed points lying in $X$. This procedure is clearly independent of the choice of compactification.

\subsection{1 dimensional subschemes} \label{1dimss}

We now want to apply Theorem~\ref{ord_loc} to the integral in eq.~(\ref{Z_cm_factor}) over $[I_{n_\beta}(X,\beta)]$. For $X$ a smooth projective surface, this class exists as the corresponding Hilbert scheme of curves is smooth, as argued in Section \ref{hcofc}. For $X$ non-compact, we will again consider a toric compactification $\bar{X}$ of $X$, as illustrated in Figure~\ref{toric_comp}. This compactification is obtained by gluing in a set of torically invariant divisors which have vanishing intersection with the compactly supported divisors of $X$. For $\beta$ the class of such a divisor, it follows that 
\be
I_{n_\beta}(X,\beta) \cong I_{n_\beta}(\bar{X},\beta)  \,,
\ee
as for $D$ such that $[D]=\beta$, all divisors linearly equivalent to $D$ will lie within $X$. We conclude that for $\beta$ the class of a compactly supported divisor, $I_{n_\beta}(X,\beta)$ is smooth on a smooth quasi-projective toric surface as well.

\subsubsection{The toric fixed points}

Above, we considered ideal sheaves corresponding to 0 dimensional subschemes. The central property in that analysis, that ideal sheaves invariant under the
torus action are monomial, i.e. locally generated by monomials, holds for ideal sheaves of any dimension. We will describe such an ideal sheaf $\cI$ by specifying it locally on
the torically invariant open sets $U_i$ of the surface $X$,
$I_i=\cI(U_i) \subset \IC[x,y]$, such that restrictions to overlaps
coincide. The monomial ideals $I_i$ are in a one-to-one relation to Young tableaux which in distinction to the 0 dimensional case may be infinite, i.e. the generators do not necessarily include monomials of the form $x^m$ or $y^n$.

\begin{figure}[h]
 \centering
  \includegraphics[width=4cm]{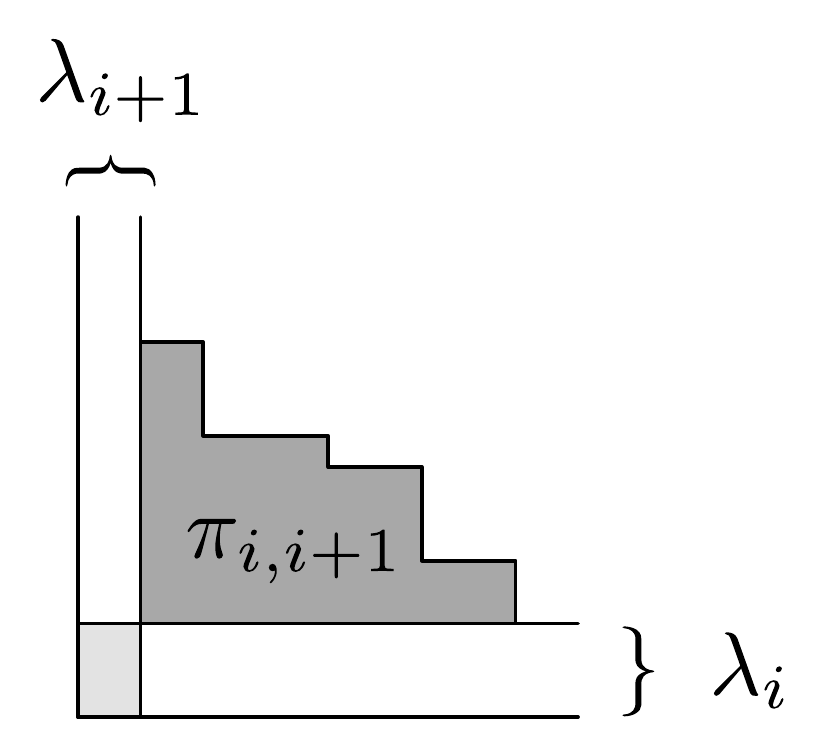}
 \caption{\footnotesize{Decomposition of a subscheme into a reduced and a 0 dimensional component.}}
 \label{crystal}
\end{figure}

The factorization (\ref{Hilbertfact}) of the Hilbert scheme into a divisorial and a punctual part is immediate when restricting to the toric fixed points: the possible associated primes to $I_i$ are $(x)$, $(y)$, and $(x,y)$ (see e.g. ref.~\cite{EisenbudHarris} for an explanation of this notion), the latter implying the existence of an embedded point. It is easy to see that all Young diagrams other than hook diagrams correspond to closed subschemes with embedded points. The decomposition 
\be
\{\mbox{infinite Young tableau}\} ~ \longleftrightarrow~ \big(\IN
\cup \{0\} \big)^2 \times \{\mbox{finite Young tableau}\}
\ee
illustrated in Figure~\ref{crystal} hence corresponds to the decomposition of the fixed point into an effective divisor, and free and embedded point contributions (free in case $\lambda_i=\lambda_{i+1}=0$).

We now consider the two factors contributing to $Z_{cm}$ as given in eq.~(\ref{Z_cm_factor}) in turn.

\paragraph{Points.}  The torically invariant ideal sheaves of points
were discussed in Section~\ref{0dimss}. They are in one-to-one
correspondence with tuples of finite Young tableaux $\pi_i$, one tableau assigned to each toric fixed point of the
surface, which we recall correspond to 2-cones of the
toric fan. The Euler characteristic of the corresponding subscheme
$Y_0$ is given by
\be
\chi(\cO_{Y_0}) = \sum \, |\pi_i| \,.
\ee
Below we will identify toric fixed points by the two bounding 1-cones, and whence use the notation $\pi_{i,i+1}$ for the corresponding Young tableau, as e.g. in Figure~\ref{crystal}.

\paragraph{Divisors.} As reviewed in Appendix~\ref{HJ_surfaces}, the torically invariant divisors are in
one-to-one correspondence with the 1-cones of the toric fan. Labeling the compactly supported torically invariant divisors as $D_i$, $i=1, \ldots, n$ (i.e. disregarding the two outermost 1-cones in the case of non-compact surfaces), a general effective divisor $D$ kept fixed by the toric action is parametrized by $n$ non-negative integers $\lambda_i$,
\be
D=\sum_{i=1}^n \lambda_i\, D_i \,.
\ee
The Euler characteristic of $D$ is then computed via (\ref{chioy_eval}),
\be
\chi(\cO_{D}) = \sum_{i=1}^n\, \Big(a_i\, \frac{\lambda_i \,
  (\lambda_i-1)}{2} + \lambda_i - \lambda_i\, \lambda_{i+1} \Big) \,.
\ee
The self-intersection numbers are here denoted $D_i^2 = - a_i$. We have furthermore set $\lambda_{n+1}=\lambda_1$ in the compact case and
$\lambda_{n+1}= 0$ in the non-compact case. We discuss how to determine the intersection matrix of the compactly supported prime divisors in Appendix~\ref{HJ_surfaces}.

\subsubsection{Crystal melting}

To emphasize the similarity with melting crystal combinatorics in six
dimensions~\cite{ORV}, we can express the Euler characteristic
$\chi(\cO_Y)=\chi(\cO_{Y_0})+\chi(\cO_{D})$ in terms of the infinite Young tableaux $\tilde{\pi}_{i,i+1}$ as
\be
\chi(\cO_{Y}) = \sum_{i=0}^{n}\, |\tilde{\pi}_{i,i+1}| +
\sum_{i=1}^n\, \Big(a_i\, \frac{\lambda_i\, (\lambda_i-1)}{2} +
\lambda_i \Big) \,,
\ee
where the box count of the infinite Young tableaux is defined as (see Figure~\ref{crystal})
\ba
|\tilde{\pi}_{i,i+1}| &:=&  \Big(\, \sum_{(I,J) \in \tilde{\pi}_{i,i+1}
  \cap [0,1, \ldots,N]^2} \, 1 \, \Big) - (N+1) \, \lambda_i - (N+1)\,
\lambda_{i+1}  \ , \hspace{1cm} N \gg 0 \\[4pt]
&=& \Big(\, \sum_{(I,J) \in \pi_{i,i+1} }\, 1 \, \Big) - \lambda_i\, \lambda_{i+1} \,.
\ea
This determines the Boltzmann weight for dissolved atoms in a crystal
described by the combinatorial quantities $\tilde\pi_{i,i+1}$ and
$\lambda_i$. Note that for the Calabi-Yau case, the self-intersection numbers are
all given by $a_i = 2$ and the Euler characteristic simplifies to
\be
\chi(\cO_{Y}) = \sum_{i=0}^{n}\, |\tilde{\pi}_{i,i+1}| + \sum_{i=1}^n\, \lambda_i^2 \,.
\ee

\subsection{The vertex formalism for toric surfaces}
We can now easily summarize the computation of the partition function
$Z_{cm}(X;q,w)$ in terms of a simple set of vertex rules:

\begin{enumerate}
\item Draw the dual web diagram of the toric fan. 2-cones are dual to
  vertices, and 1-cones are dual to legs.
\item Each vertex $i$ carries two positive integer labels $\lambda_i$ and
  $\lambda_{i+1}$ (``one-dimensional Young tableaux''), one assigned to
  each emanating leg, and contributes a vertex factor
\be
V_{\lambda_i, \lambda_{i+1}}(q) = \frac{1}{\hat\eta(q)} ~ q^{-\lambda_i \,\lambda_{i+1}}
\ee
to the partition function.
\item Vertices are glued along legs carrying the same integer label
  $\lambda_i$ with a gluing factor 
\be
G_{\lambda_i}(q,w_i) = q^{a_i \, \frac{\lambda_i
    \,(\lambda_i-1)}{2} + \lambda_i} \, w_i^{\lambda_i} \,,
\ee
where the self-intersection numbers $-a_i$ are determined graphically as described below. $w_i$ labels the homology class of the curve corresponding to the leg along which the vertices are glued. 
\item Multiplying the vertex and gluing factors together, summing 
  the $\lambda_i$ on internal legs over all non-negative integers while setting those on external legs to zero then yields the
  melting crystal
  partition function $Z_{cm}(X;q,w)$.
\end{enumerate}

\bigskip
\bigskip

\begin{figure}[h]
 \centering
  \includegraphics[width=8cm]{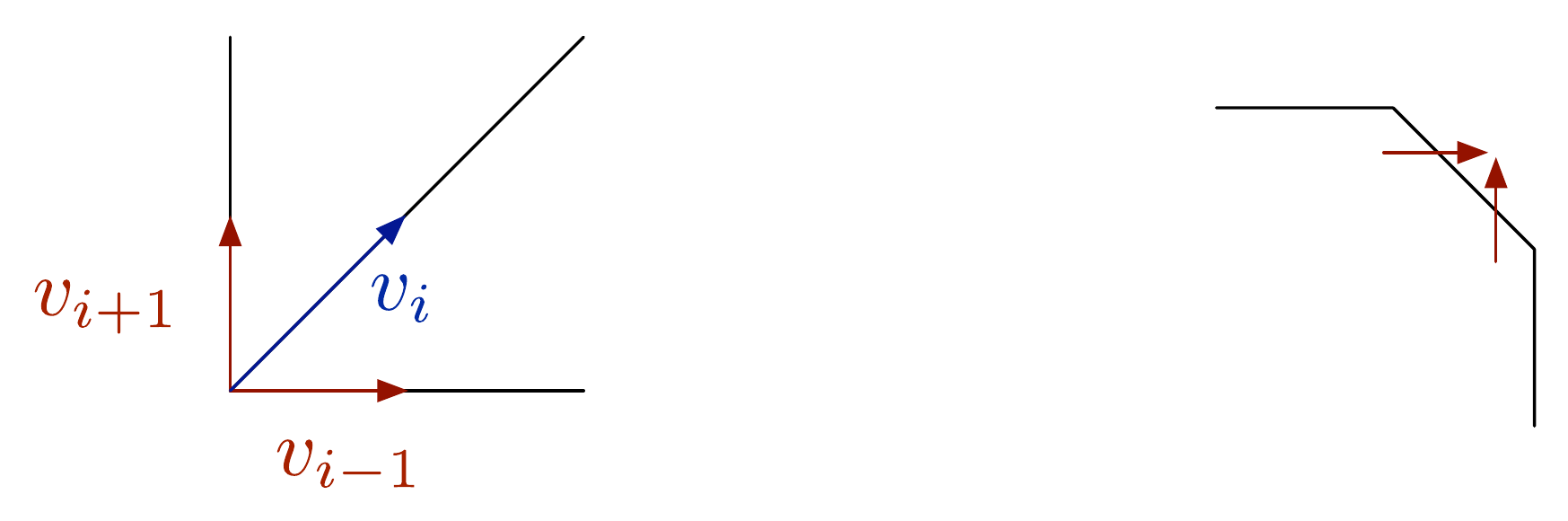}
 \caption{\footnotesize{An example of a curve with self-intersection
     number $-1$ (the exceptional divisor of the blow-up of $\IC^2$
     at the origin). On the right-hand side, we have indicated the
     dual web diagram with the analogue of the framing vectors of the six-dimensional vertex formalism~\cite{AKMV,LLLZ}.}}
 \label{self_intersection}
\end{figure}

We can determine the self-intersection numbers $-a_i$ graphically as follows. Recall that they are given by the relation
\be
a_i \, v_i = v_{i-1} + v_{i+1}
\ee
between the generator $v_i$ of the 1-cone associated to $D_i$ and
those of the two neighboring 1-cones, see Figure~\ref{self_intersection}.
Note that since the surface $X$ is non-singular by assumption, one has 
\be
v_{i-1} \times v_i = v_{i} \times v_{i+1}=1 \,,
\ee
where the $\times$ product computes the volume of the cell spanned by the generators. Hence
\be
a_i = v_{i-1} \times v_{i+1} \,.
\ee

\subsection{Examples} \label{examples}

\subsubsection{Projective plane}

\begin{figure}[h]
 \centering
  \includegraphics[width=5cm]{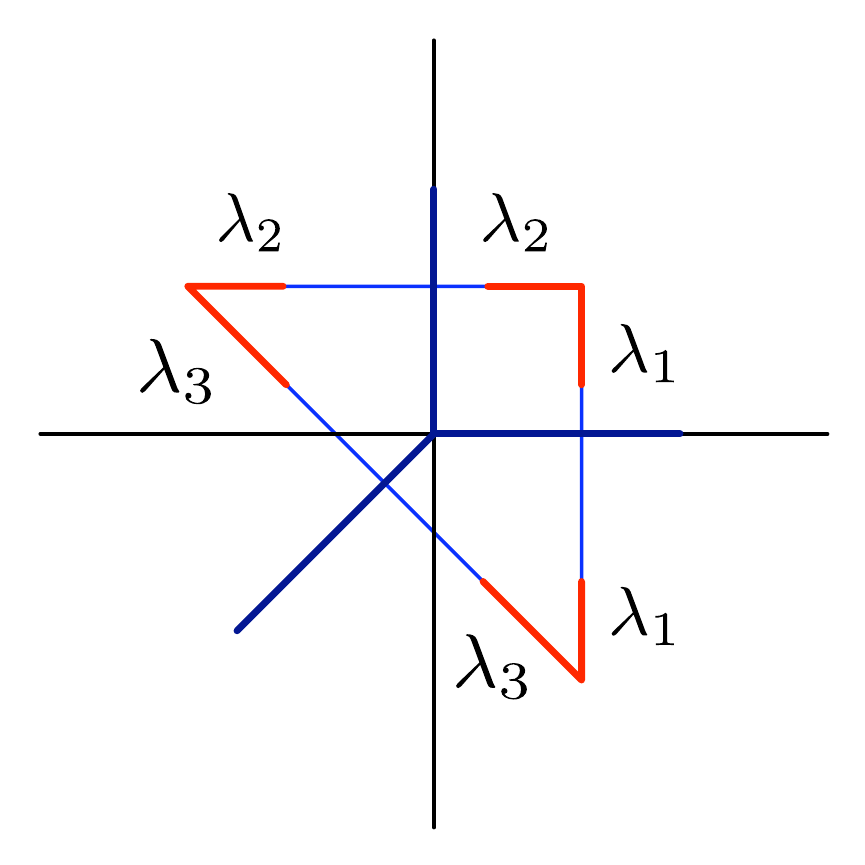}
 \caption{\footnotesize{The toric fan for $\IC \IP^2$, and the corresponding web diagram, with the legs of the vertices labelled.}}
 \label{P_2}
\end{figure}

The toric fan and web diagram of the projective space $\IC \IP^2$ are
depicted in Figure~\ref{P_2}. The self-intersection numbers $-a_i$ of the torically invariant
divisors are $a_i=-1$. The partition function is thus
\ba
Z_{cm}(\IC \IP^2;q,w) &=& \sum_{\lambda_1, \lambda_2, \lambda_3 =
  0}^{\infty}\, \frac{1}{\hat\eta(q)}\, q^{-\lambda_1 \, \lambda_2}~
q^{-\frac{\lambda_2^2}{2}+\frac{3}{2}\,\lambda_2} \,w^{\lambda_2} ~
\frac{1}{\hat\eta(q)}\, q^{-\lambda_2\, \lambda_3} ~
q^{-\frac{\lambda_3^2}{2}+\frac{3}{2}\, \lambda_3} \, w^{\lambda_3}
\\ && \hspace{2cm} \times\, \frac{1}{\hat\eta(q)}\, q^{-\lambda_3 \, \lambda_1} ~
q^{-\frac{\lambda_1^2}{2}+\frac{3}{2}\, \lambda_1} \, w^{\lambda_1} \\[4pt]
&=& \frac{1}{\hat\eta(q)^3} \, \sum_{\lambda_1, \lambda_2, \lambda_3 =
  0}^{\infty} \, q^{-\frac{1}{2}\, (\lambda_1 + \lambda_2 +
  \lambda_3)^2 + \frac{3}{2}\, (\lambda_1 + \lambda_2 + \lambda_3)} \,
w^{\lambda_1+\lambda_2+\lambda_3} \ ,
\ea
with $w$ labeling the hyperplane class.

As a check, we can use this formula to extract the Euler characteristic $\chi\left(I_{n_\beta}(X,d)\right)$ of the moduli space of degree $d$ divisorial curves on $X=\IC \IP^2$. By our result for the partition function, it is given by the number of ways to obtain $d$ as the sum of three non-negative integers,
\be
d = \lambda_1 + \lambda_2 + \lambda_3 \,.
\ee
As $|I_{n_\beta}(X,d)| = \IP(H^0(X,\cO_X(D))$ for a choice of divisor $D$ with $[D] = \beta$, (see e.g. \cite[p.~137]{GriffithsHarris}), and $\chi(\IC\IP^n) =n+1$, this is indeed the correct result.

\subsubsection{Hirzebruch-Jung surfaces\label{ALEexamples}}
\begin{figure}[h]
 \centering
  \includegraphics[width=12cm]{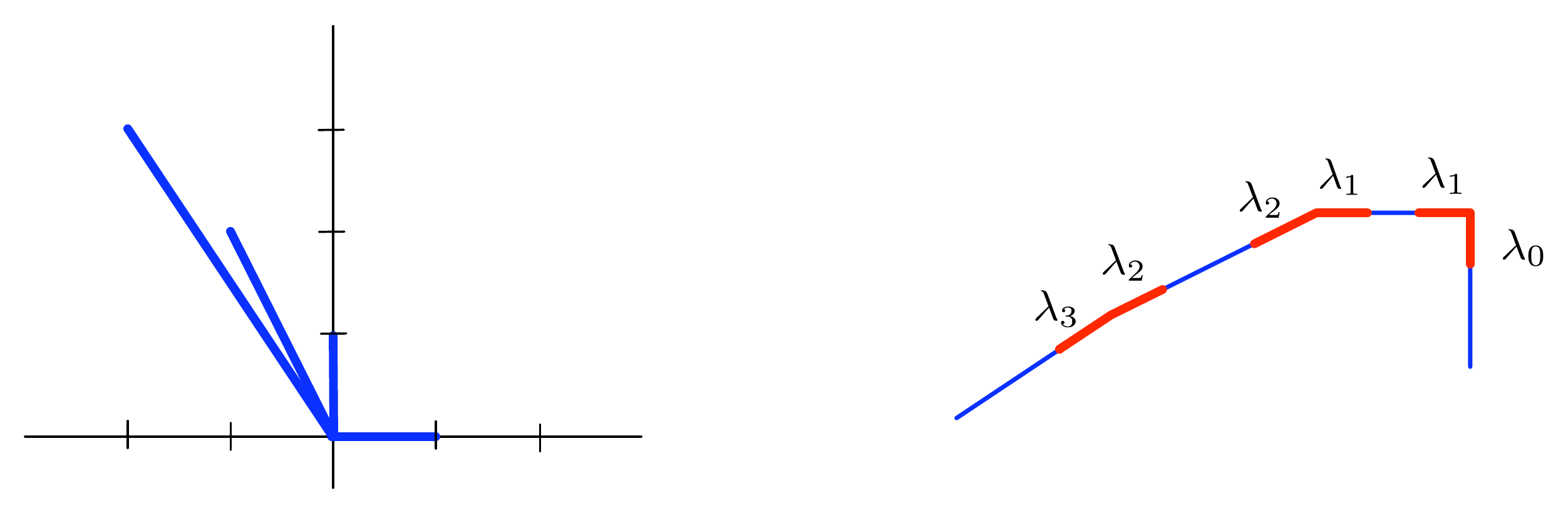}
 \caption{\footnotesize{The toric fan for $A_2$, and the corresponding web diagram, with the legs of the vertices labelled.}}
 \label{an}
\end{figure}

A choice of 1-cones describing the Hirzebruch-Jung surfaces $Y_{p,q}$ is given in Appendix \ref{HJ_surfaces}. The corresponding fan for the example $Y_{3,2}=A_2$ is depicted in Figure~\ref{an}, together with the dual web diagram. 

The vertex rules yield the partition function
\ba
Z_{cm}(Y_{p,q};q,w) &=& \sum_{\lambda_1, \ldots, \lambda_n = 0}^{\infty}\,
\frac{1}{\hat\eta(q)}\, q^{-\lambda_0\, \lambda_1} ~ q^{\frac{1}{2}a_1\lambda_1^2+\lambda_1(1-\frac{1}{2}a_i)}
\, w_1^{\lambda_1} ~ \frac{1}{\hat\eta(q)}\, q^{-\lambda_1\,
  \lambda_2} ~ \cdots ~ \\
&& \hspace{2cm} \times  q^{\frac{1}{2}a_n \lambda_n^2+\lambda_n(1-\frac{1}{2}a_n)} \, w_n^{\lambda_n} ~
\frac{1}{\hat\eta(q)}\, q^{-\lambda_n \, \lambda_{n+1}} \\[4pt]
&=& \frac{1}{\hat\eta(q)^{n+1}} \, \sum_{\lambda_1, \ldots, \lambda_n = 0}^{\infty}\, q^{\frac{1}{2}\lambda \cdot C \lambda - \frac{1}{2} \lambda \cdot Ce - \frac{1}{2}\lambda_1 - \frac{1}{2}\lambda_n} \,w^{\lambda} \,,
\ea
where we have defined $e:=(1, \dots, 1)$, $\lambda:=(\lambda_1,\dots, \lambda_n)$, and $w^\lambda:=w_1^{\lambda_1}\cdots w_n^{\lambda_n}$. The negative of $C$ is the
intersection matrix (\ref{int_mat}) of the compact divisors given in
Appendix~\ref{HJ_surfaces}, where we also review how to determine the self-intersection numbers $-a_i$ for these surfaces.

ALE spaces have vanishing canonical class. By specializing the above formula to this case, with all $a_i=2$, we observe the ensuing simplification to
\ba
Z_{cm}(A_n;q,w) &=& \frac{1}{\hat\eta(q)^{n+1}} \, \sum_{\lambda_1, \ldots, \lambda_n = 0}^{\infty}\, q^{\frac{1}{2}\lambda \cdot C \lambda} \,w^{\lambda} \,.
\ea

\section{Gauge theory} \label{gauge}

On non-compact surfaces, the vertex rules from the previous section
only capture part of the complete gauge theory partition function. To
obtain the full partition function, we need to include contributions
from both negative and non-compact divisors. In the compact case, linear equivalence will furthermore identify divisors in the gauge theory that correspond to distinct fixed points of the torus action, changing the combinatorics of the problem.

The factorization (\ref{torsion_free}) of rank 1 torsion-free sheaves $\cT$,
\be
\cT = \cL \otimes \cI_Z    \,,
\ee
implies that the moduli space of isomorphism classes of torsion-free sheaves 
factorizes
$$
\fM_X(\beta,n)= \pic_\beta^X \times X^{[n-n_\beta]} \, ,
$$
with the Picard group $\pic_\beta^X$ parametrizing line bundles which contribute $n_\beta =- \frac{1}{2} \beta\cdot(\beta +K_X)$ to the Euler characteristic of the torsion-free sheaf. It follows
that the generating function (\ref{Zgt}) for the counting problem splits into a discrete and a continuous part,
\be
Z_{gt}(X;q,w)=\sum_{\beta\in
  H_2(X,\IZ)} \sum_{\cL\in \pic_\beta^X} \, q^{-n_\beta}\,w^\beta~\sum_{n\geq 0}~\int_{X^{[n]}}\,
e\big(TX^{[n]}\big)~q^n \ .
\ee

For the continuous part, we need to count 0 dimensional subschemes. These contribute identically to $Z_{gt}$ and $Z_{cm}$, by the factor
\be
\frac{1}{\hat\eta(q)^{\chi(X)}} \,,
\ee
as determined in Section ~\ref{0dimss}.

It remains to enumerate the holomorphic line bundles $\cL \in \pic^X_\beta$. In fact, on a toric manifold, the Picard group is spanned by the classes of torically invariant divisors. Our task will be to determine an integral generating set among these. We do this for each of the examples considered in Section~\ref{examples} in turn.\footnote{Note the deceptive similarities to the case of counting subschemes. The sum over torically invariant divisors there was due to localization. Furthermore, we will start off here by considering two additional toric divisors (those of non-compact support), but taking linear equivalence into account will result in the same number of summations as previously.}

\subsection{Projective plane}

The homology of $\IC \IP^2$ is spanned by the hyperplane class, with
self-intersection number~1. Holomorphic line bundles on $\IC\IP^2$ hence permit self-dual, but not anti-self-dual connections. Of course, the two conditions are interchanged upon reversing the orientation of the surface. Let us proceed to determine the gauge theory partition function, in the sense developed in Section~\ref{instantons}, upon replacing anti-self-duality by self-duality. 

The three torically invariant divisors of the complex projective plane
are linearly equivalent (the Picard group of complex
projective space in any dimension is spanned by the hyperplane
divisor). In contrast to the non-compact examples discussed below, we obtain the
gauge theory partition function from the melting crystal partition
function simply by dropping the sum over equivalent bundles and the
restriction to effective divisors, and by taking into account the
change in weight due to $K_{\IC \IP^2}\neq0$. One thereby finds
\be 
Z_{gt}(\IC \IP^2;q,w) = \frac{1}{\hat\eta(q)^3} \, \sum_{u=-\infty}^{\infty}\, q^{-\frac{1}{2}\, u^2 }\, w^u  \,.
\ee

\subsection{Hirzebruch-Jung surfaces}

When including non-compact prime divisors, the full set of divisors
associated to the 1-cones of the toric fan of a $Y_{p,q}$ space become linearly
dependent. We will now determine an integral generating set for the
Picard group. For clarity, we will first treat the simpler case of ALE
spaces $A_n$, though they are of course encompassed by the subsequent treatment
of general Hirzebruch-Jung surfaces $Y_{p,q}$, upon setting $(p,q)=(n+1,n)$.

\paragraph{ALE spaces.} \ 
Consider the vectors $(1, 0)$ and $(0, 1)$ in the toric fan of the
resolved geometry of $A_n = \IC^2/\IZ_{n+1}$ introduced in Appendix~\ref{HJ_for_real}. They correspond to the two principal
divisors\footnote{See e.g. ref.~\cite{Fulton} for the notation $\chi^u$, which assigns a function to the lattice vector $u$.}
\ban
\Div\big(\chi^{(1, 0)}\big) &=& D_0 - D_2 -2 \,D_3 - \ldots -n\, D_{n+1} \,, \nn \\[4pt]
\Div\big(\chi^{(0 ,1)}\big) &=& D_1 + 2\, D_2 + \ldots + (n+1)\, D_{n+1} \,,  \label{lin_dep}
\ean
where we have labelled toric divisors in anti-clockwise order. The two divisors corresponding to the outermost 1-cones, $D_0$ and $D_{n+1}$, have non-compact support. Based on the
relations of linear equivalence induced by eq.~(\ref{lin_dep}), we now
demonstrate that the classes
\ben
e^i:=-\sum_{j=1}^{n}\, \big(C^{-1}\big)^{ij} \, [D_j]  \,\,,\quad\quad i=1, \ldots,n \,, \label{def_basis}
\een
with $-C$ the intersection matrix of the compact divisors as given in eq.~(\ref{int_mat}) of Appendix~\ref{HJ_surfaces}, constitute an integral generating set for the Picard group
$A_{1}(X)$.\footnote{Such a generating set is of course not
  unique. Our choice provides a dual set, via the intersection product
  linearly extended to non-compact divisors, to the compactly
  supported divisors $D_1, \ldots, D_{n}$, and as such corresponds to
  the basis of bundles constructed by Kronheimer and Nakajima in
  ref.~\cite{KronheimerNakajima}.} As the entries of $C^{-1}$ are
fractional, we need to demonstrate both that the elements $e^i$ are
generators, and that they are integral linear combinations of the toric
divisors (including $[D_0]$ and $[D_{n+1}]$). Both properties follow
upon providing the following recursive presentation of the $e^i$ (for
$n>1$; the case $n=1$ is trivial, with a single generator
$[D_0]=[D_2]$). It is easy to verify that $e^1 = [D_0]$ and $e^{n}=[D_{n+1}]$. For $i=2, \ldots, n-1$, the $e^i$ satisfy
\be
e^i = e^{i-1}-e^{n} - [D_{i}]-\ldots -[D_{n+1}] \,.
\ee
It follows that $\{e^i\}$ represents an extension of the set of non-compact torically invariant divisors $\{[D_0], [D_{n+1}] \}$ to an integral generating set for the Picard group.

Parametrizing the class of a divisor $D=D_u$ in terms of the generators $e^i$,
$$[D_u]=\sum_{i=1}^{n}\, u_i\, e^i$$ with $u=(u_1,\dots,u_n)\in\IZ^n$,
we can now compute its second Chern character. Note that the Chern classes corresponding to divisor classes are integral, irrespective of the
support of the divisor. At the level of Chern classes, we can hence
invoke the presentation of $e^1=[D_0]$ and $e^{n}=[D_{n+1}]$ in
eq.~(\ref{def_basis}) to solve for $e^1$ and $e^n$, and the right-hand side, despite the appearance
of the fraction $\frac{1}{n+1}$, maps into integral
cohomology. With the intersection pairing (\ref{instno}), we thus arrive at
\ben \label{gaugeweight}
\ch_2\big(\cO_X(D_u)\big) = \mbox{$\frac{1}{2}$}\, u\cdot C^{-1} u \,.
\een

\paragraph{General $\mbf{Y_{p,q}}$ spaces.} \ 
With the parametrization of the toric fan given in Appendix~\ref{HJ_surfaces}, the principal divisors corresponding to the lattice vectors $(1,0)$ and $(0,1)$ are
\ba
\Div \big(\chi^{(1,0)}\big) &=&  \sum_{i=0}^{n+1}\, x_i\, D_i \,, \\[4pt]
\Div \big(\chi^{(0,1)}\big) &=& \sum_{i=0}^{n+1}\, y_i\, D_i \,,
\ea
where we have introduced the notation $v_i = (x_i, y_i)$ for the generator of the $i^{th}$ 1-cone. Recalling that $v_0=(1,0)$, we arrive at the linear equivalences
\ba
[D_0] &=& \frac{1}{y_{n+1}}\, \sum_{i=1}^{n}\, (x_{n+1}\, y_i - x_i \, y_{n+1})\, [D_i] \,,\\[4pt]
[D_{n+1}] &=& - \frac{1}{y_{n+1}}\, \sum_{i=1}^{n}\, y_i \,[D_i] \,.
\ea 
By invoking eq.~(\ref{latrel}) from
Appendix~\ref{HJ_surfaces} and the relation
\be
x_{i-1}\, y_{i+1} - x_{i+1}\, y_{i-1} = a_i \,,
\ee
which follows from eq.~(\ref{latrel}) and the fact that the resolved
geometry is non-singular (i.e. the 2-cones have volume 1), we can
easily verify that in $A_{1}(Y_{p,q})\otimes \IQ$ one has
\ban
[D_0] &=& -\sum_{i=1}^{n}\, \big(C^{-1}\big)^{1 \, i}\, [D_i] \,, \\[4pt]
[D_{n+1}] &=& -\sum_{i=1}^{n}\,\big(C^{-1}\big)^{n \, i} \,[D_i] \,.  \label{lin_equ_apq}
\ean
We can now complete the set $\{[D_0],[D_{n+1}]\}$ to an integral generating set for $A_{1}(Y_{p,q})$
by setting $e^1 = [D_0]$, $e^{n}=[D_{n+1}]$ and defining $e^i$, $i=2, \ldots, n-1$ recursively via
\be
e^i = e^{i-1} - \sum_{j=i}^{n}\, c^i_j\, [D_i] - c^i_{n+1}\, e^{n} \,,
\ee
where 
\ba
c^i_i &=& 1 \,,\\[4pt]
c^i_{j} &=& a_{j-1}\, c^i_{j-1} -1 \,,\quad j=i+1, \ldots, n \,,\\[4pt]
c^i_{n+1} &=& - c^i_{n-1} + a_{n}\, c^i_{n} \,.
\ea
The generators so defined satisfy
\be
e^i = - \sum_{j=1}^{n}\, \big(C^{-1}\big)^{ij}\, [D_j] \,.
\ee
The computation now proceeds as above, yielding, for 
\be
[D_u]=\sum_{i=1}^{n}\, u_i\, e^i\,,
\ee
the second Chern character as given in (\ref{gaugeweight}).

\vspace{1.5cm}

Combining the contributions from the line bundles and the ideal sheaves of points, we obtain the partition function
\be
Z_{gt}(Y_{p,q};q,v) = \frac{1}{\hat\eta(q)^{\chi(Y_{p,q})}}\, \sum_{u \in \IZ^{n}}\, q^{-\frac{1}{2}\, u\cdot C^{-1} u} \,w^u
\ee
with $w^u:=w_1^{u_1}\cdots w_{n}^{u_{n}}$. This coincides with the results obtained in ref.~\cite{FMP}.

\subsection*{Acknowledgements}

We would like to thank Ugo Bruzzo, Francesco Fucito, Elizabeth Gasparim, Antony Maciocia, Jose F. Morales, Rubik Poghossian, Alessandro Tanzini and Constantin Teleman for discussions. AK would like to thank Pierre Cartier, Ofer Gabber, Nicol\`o Sibilla, and especially Ilya Shapiro for patient explanations. MC is partially supported by the Funda\c{c}\~{a}o para a Ci\^{e}ncia e a
Tecnologia (FCT/Portugal), and by the Center for Mathematical Analysis, Geometry and Dynamical Systems. AK is supported in part by {\it l'Agence  
Nationale de la Recherche} under the grant ANR-BLAN06-3-137168. RJS is partially supported by grant ST/G000514/1 ``String Theory Scotland'' from the UK Science and Technology Facilities Council.

\appendix

\section{Euler characteristic of torus invariant subschemes}
\label{app_cech}

In this appendix, we will calculate the Euler characteristic $\chi(\cO_Y)$ of torically invariant subschemes $Y$ of a non-compact toric surface $X$ using \v{C}ech
cohomology. We will compute the cohomology with respect to the canonical
torically invariant open cover $\{U_i\}$ of $X$, where each $U_i$
corresponds to a maximal cone. We choose the index $i=0,\ldots,n$ to enumerate consecutive maximal cones in anti-clockwise order. The collection of sets thus defined has the properties
\ben \label{prop_open}
U_i \cap U_j ~
\begin{cases} 
  ~ = ~ (\IC^*)^2 \ , & \text{$j \neq i \pm 1$} \ ,\\ 
  ~ \supset ~ (\IC^*)^2 \ , &\text{$j= i\pm 1$} \ ,
\end{cases} 
\een
and
\be
U_i \cap U_j \cap U_k = (\IC^*)^2
\ee
for any $i,j,k$. Our strategy to compute $\chi(\cO_Y)$ is as follows. Given 
\ben
V=\bigcup_{i=a}^b\, U_i\,,   \label{V}
\een
let
\be
A_V = \cO_Y(U_{a-1} \cup V \cup U_{b+1})\big|_V 
\ee
be the space of global sections of $\cO_Y(V)$ that lift to
$\cO_Y(U_{a-1})$ and $\cO_Y(U_{b+1})$, and define
\be
\chi_V = \dim_\IC(A_V) -  \check{h}^1\big(\cO_Y\big|_{U_{a-1} \cup V \cup U_{b+1}}\big) \,.
\ee
Note that we avoid the use of $\check{h}^0(\cO_Y|_{U_{a-1} \cup V \cup
  U_{b+1}})$ in place of $\dim_\IC(A_V)$, as the former could be infinite. We will compute $\chi_V$ for $V= U_i$, and given two adjacent such sets, such as $V$ in eq.~(\ref{V}) and $$W=\bigcup_{i=b+1}^c\, U_i \,,$$
together with the integers $\chi_V$ and $\chi_W$, we will determine $\chi_{V\cup W}$. Applying this gluing operation a finite number of times will yield $\chi(\cO_Y)$.

The monomial generators of $\cO_Y(U_i)$ are, as explained in Section~\ref{0dimss}, in one-to-one
correspondence with the boxes of a possibly infinite Young tableau
$\pi$, such that the number of boxes in the $k$-th row and column of $\pi$ stabilize at large $k$. We will denote these stable values as $\lambda_{i+1}$ and $\lambda_{i}$, respectively. The boxes with coordinates $(k,l)$, $k \ge \lambda_i$, $l \ge \lambda_{i+1}$ correspond to sections that restrict to 0 outside of $U_i$, and therefore can be attributed to the punctual factor in the decomposition (\ref{Hilbertfact}). Each such box contributes a summand of 1 to $\chi(\cO_Y)$. In the following, we can hence restrict attention to subschemes $Y$ without such free or embedded points, corresponding to hook Young tableaux. 

We turn to the calculation of $\chi_{U_i}$. We can choose coordinates in the three neighboring patches $U_{i-1}, U_{i}, U_{i+1}$ such that
\ba
\cO_Y(U_{i-1}) &=& \IC\big[\mbox{$\frac{1}{x}$}\,,\, x^{a_{i}}
\,y\big] \, \big/\,
\big((\mbox{$\frac{1}{x}$})^{\lambda_{i-1}}\,(x^{a_{i}}
y)^{\lambda_i} \big) \ , \\[4pt]
\cO_Y(U_i) &=& \IC[x,y] \,\big/\, \big(x^{\lambda_{i+1}}\,
y^{\lambda_{i}}\big) \ , \\[4pt]
\cO_Y(U_{i+1}) &=& \IC\big[x\, y^{a_{i+1}}\,,\, \mbox{$\frac{1}{y}$}\big]
\,\big/\, \big((x \,y^{a_{i+1}})^{\lambda_{i+1}}\,
(\mbox{$\frac{1}{y}$})^{\lambda_{i+2}}\big) \,,
\ea
together with
\ba
\cO_Y(U_{i-1,i}) &=& \IC\big[x\,,\,\mbox{$\frac{1}{x}$}\,,\, y\big]
\,\big/\, \big(y^{\lambda_{i}}\big) \ , \\[4pt]
\cO_Y(U_{i,i+1}) &=& \IC\big[x\,,\,y\,,\,\mbox{$\frac{1}{y}$}\big]
  \,\big/\, \big(x^{\lambda_{i+1}}\big) \,,
\ea
and
\be
\cO_Y(U_{i-1,i,i+1}) = 0 \,.
\ee
We have introduced the notation $U_{i,\ldots ,j}=U_i \cap \ldots \cap U_j$. The negative self-intersection numbers $a_i$ are discussed in Section~\ref{vertex} and Appendix~\ref{HJ_surfaces}. A moment's thought yields
\ba
\dim_\IC(A_{U_i}) &=& \sum_{s=0}^{\lambda_i-1} ~\sum_{r=0}^{a_{i}\,s} \,1 + \sum_{s=0}^{\lambda_{i+1}-1}~ \sum_{r=0}^{a_{i+1}\, s}\, 1 - \lambda_{i+1}\,\lambda_i \\[4pt]
&=& a_{i}\, \frac{(\lambda_i-1)\, \lambda_i}{2} + \lambda_i + a_{i+1}\,
\frac{(\lambda_{i+1}-1)\, \lambda_{i+1}}{2} + \lambda_{i+1} -
\lambda_{i}\, \lambda_{i+1}  \,,
\ea
and
\be
\check{H}^1\big(\cO_Y\big|_{U_{i-1} \cup U_i \cup U_{i+1}}\big) = 0  \,,
\ee
hence
\be
\chi_{U_i} = \dim_\IC(A_{U_i})  \,.
\ee
Next, we turn to the compution of $\chi_{V \cup W}$ given $\chi_V$ and
$\chi_W$, using the notation for $V$ and $W$ introduced above. Note
that by eq.~(\ref{prop_open}), $V \cap W = U_b \cap U_{b+1}$. If we
consider the sum $\chi_V + \chi_W$ as an approximation to $\chi_{V
  \cup W}$, then we make the following mistakes:
\begin{itemize}
\item We count generators of $A_V$ and $A_W$ that have the same restriction to $V \cap W$ twice.
\item We count generators of $A_{V}|_{V \cap W}$ that do not lift to $A_W$, and likewise elements of $A_{W}|_{V \cap W}$ that do not lift to $A_V$.
\item We do not subtract new contributions to $\check{H}^1$, i.e. generators in $\cO_Y(V \cap W)$ that are not exact.
\end{itemize}
Now consider the space $B_{V,W}=\cO_Y(U_b \cup U_{b+1}) |_{U_b \cap U_{b+1}}$. Generators of this space either lift to elements in both $A_V$ and $A_W$, or in either $A_V$ or $A_W$ but not both, or in neither. By the following lemma, the generators of $B_{V,W}$ are hence in one-to-one correspondence with the elements over-counted above. 
\vspace{0.5cm}

\begin{1cech}
Elements in $B_{V,W}$ that lift neither to elements in $A_V$ nor in $A_W$ lie in $\check{H}^1(V\cup W)$.
\end{1cech}
\begin{proof}
A preimage $r$ under the \v{C}ech differential $\delta$ of a monomial element in $\check{C}^1(V \cup W)$ of the form 
\be
s_{ij} =
\begin{cases}
a & \text{if $\{i,j\} = \{b,b+1\}$ \ ,} \\
0 & \text{otherwise}
\end{cases}
\ee
must satisfy 
\be
r_i-r_j \big|_{U_{i,j}} =
\begin{cases}
\pm \, a & \text{if $\{i,j\} = \{b,b+1\}$ \ ,} \\
0 & \text{otherwise \ .}
\end{cases}
\ee
Such an element exists if and only if $a$ lifts to an element in $A_V$ or $A_W$.
\end{proof}
Finally, the number of generators of $B_{V,W}$ already entered into our computation of the dimension of $A_{U_i}$ above,
\be
\dim_\IC(B_{V,W}) =  \sum_{s=0}^{\lambda_{b+1}-1}~ \sum_{r=0}^{a_{b+1}\, s}\, 1 \,.
\ee
Combining all of these observations, we arrive at the desired formula
for the holomorphic Euler characteristic of $Y$,
\be
\chi(\cO_{Y}) = \sum_{i=1}^n\, \Big(a_i\, \frac{\lambda_i\, (\lambda_i-1)}{2} + \lambda_i - \lambda_i\, \lambda_{i+1}\Big) \,.
\ee 
Note that with very little extra effort, this calculation can be modified to encompass the case of compact $X$ as well.

\section{Toric surfaces} \label{HJ_surfaces}

\subsection{General non-singular toric surfaces}
A non-singular toric surface is determined by a sequence of integral vectors $v_i$ in $\IZ^2$, taken in counter-clockwise order, that generate the 1-cones of the toric fan. For compact surfaces, we will enumerate these $v_1$ through $v_n$. Any two adjacent vectors span a 2-cone in this case, and by non-singularity, generate the lattice. For non-compact surfaces, two of the 2-cones have only one neighbor. For this case, we denote the outer-most vectors by $v_0$ and $v_{n+1}$, giving rise to a total of $n+2$ 1-cones.

The coordinate transformation between neighboring 2-cones is of the form
$$(x,y) ~ \longmapsto~ \big(\mbox{$\frac{1}{x}$}\,,\, x^a\,
y\big)\,.$$ For each 2-cone, generated by integral vectors $v_i$ and
$v_{i+1}$, one determines the integer $a=a_i$ by considering the generator $v_{i+1}$ of the neighboring cone, counter-clockwise, which satisfies
\ben
v_{i+1} = -v_{i-1} + a_i\, v_i \,.  \label{intersection_n} 
\een

The torically invariant prime divisors of a toric manifold are in
one-to-one correspondence with the integral generators of the 1-cones. In the conventions introduced above, the number of compactly supported divisors is $n$, both in the compact and the non-compact case. The intersection matrix of the compactly supported divisors is determined as follows. Two divisors whose associated 1-cones span a 2-cone of the fan intersect
transversally, while all others are disjoint. The negative self-intersection number of the divisor $D_i$ associated to the generator $v_i$ is given by the constant $a_i$ in eq.~(\ref{intersection_n}), $D_i^2 = - a_i$ (recall that we are excluding $i=0$ and $i=n+1$; indeed, due to the non-compact support of the associated divisors, the naive
intersection number here is not defined). The intersection matrix $-C$ for the compactly supported toric divisors therefore has the form
\ben   \label{int_mat}
C= 
\left(\begin{matrix}
a_1 & -1 & 0 & \ldots & 0 \\
-1 & a_2 & -1 & \ldots & 0 \\
0 & -1 & \ddots &  & \vdots \\ 
\vdots & \vdots &  & \ddots & -1 \\ 
0 & 0 & \ldots & -1 & a_{n}
\end{matrix}\right) \,.
\een

\subsection{Hirzebruch-Jung surfaces} \label{HJ_for_real}
Hirzebruch-Jung spaces $X=Y_{p,q}$ are non-compact toric surfaces, parametrized by two positive coprime
integers $p$ and $q$ with $p>q$. They are defined as the resolutions of $A_{p,q}$
quotients, i.e. the quotients of $\IC^2$ by the action of the cyclic
group $\IZ_p$ generated by
\be
\Gamma = \left(
\begin{matrix}
\xi & 0 \\
0 & \xi^q \\
\end{matrix}
\right) \,,
\ee
where $\xi = \e^{2 \pi \ii /p}$. The toric fan for the singular space is given by the two 1-cones generated by the integral vectors $v_0 = (1,0)$ and $v_{n+1} = (-q,p)$ respectively, and the 2-cone generated by the pair. The resolution is obtained via subdivision with 1-cones generated by integral vectors $v_1, \ldots, v_{n}$ such that
\ben
v_{i-1} + v_{i+1} = a_i\, v_i  \label{latrel}
\een
for $i=1, \ldots, n$. The integers $a_i$ can be read off from the continued fraction expansion of $p/q$,
\be
\frac{p}{q} = a_1 - \frac{1}{a_2 - \frac{1}{\ddots a_{n-1}-\frac1{a_{n}}}} \,.
\ee
With these entries, $C$ of eq.~(\ref{int_mat}) is positive definite, the intersection matrix hence negative definite.

\begin{figure}[h]
 \centering
  \includegraphics[width=5cm]{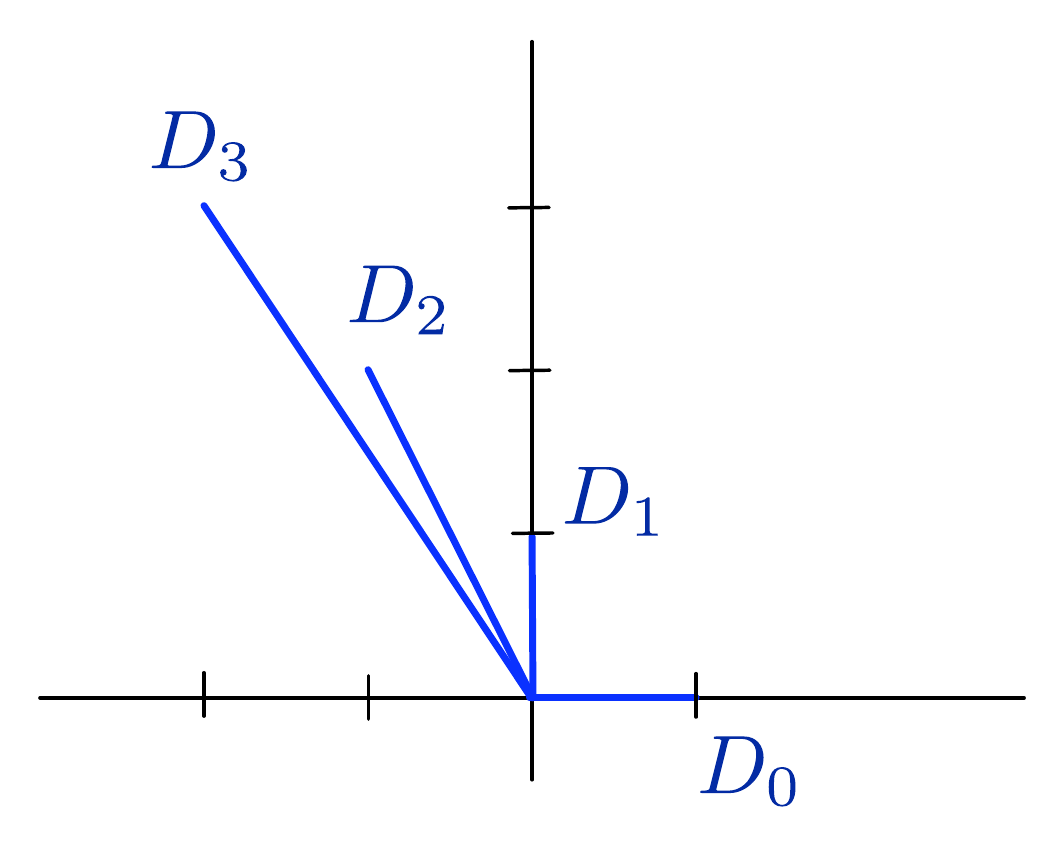}
 \caption{\footnotesize{The toric fan for $A_2$, with the torically invariant divisors indicated.}}
 \label{a2}
\end{figure}

Topologically, ALE spaces are resolutions of $A_n$ singularities. These are the Hirzebruch-Jung spaces $Y_{n+1,n}$. For these surfaces, $a_i = 2$ for $i=1 , \ldots, n$, and a choice of integral vectors generating the 1-cones of the toric fan is given by
\be
v_0 = (1,0) \,, \quad v_1 = (0,1) \,, \quad  \ldots \,, \quad v_{n+1} = (-n,n+1) \,.
\ee
We have depicted the fan for the surface $A_2$ in Figure~\ref{a2}.

\section{Characteristic classes of coherent sheaves} \label{B}

In physics, we are most familiar with characteristic classes assigned
to vector bundles on a manifold $X$, taking values in $H^*(X,\IZ)
\otimes \IQ$ (the sheaf cohomology of the locally constant sheaf); we often deal with the image in de Rham cohomology. Chern classes form basic building blocks for all characteristic classes. The total Chern class satisfies the multiplicative property
\ben
c_t (\cE) = c_t(\cE'\,) \cdot c_t(\cE''\,)   \label{multiplicativity}
\een
whenever the bundles $\cE$, $\cE'$, and $\cE''$ fit into an exact sequence
\be
\xymatrix{
0 ~ \ar[r] &~ \cE'~ \ar[r] & ~ \cE ~ \ar[r] & ~ \cE''~ \ar[r] & ~ 0 \ .
}
\ee
The product in eq.~(\ref{multiplicativity}) is the cup product or the wedge product, respectively.

A refined version of characteristic classes takes values in the Chow ring $A_*(X) \otimes \IQ$ of $X$. One path between the two definitions is via the splitting principle and the relation of line bundles to divisors. We take the image of an irreducible closed subscheme $Y$ in the Chow ring to be the underlying closed subset endowed with the reduced induced structure, $Y_{red}$, with {\it multiplicity} given by the length of the local ring $\cO_{y,Y}$ at the generic point $y$ of $Y_{red}$ (we will unravel this perhaps unfamiliar sounding definition below in the simple case of a 0 dimensional subscheme).

The Grothendieck group $K(X)$ of a scheme $X$ is the free abelian
group generated by the coherent sheaves on $X$, modulo the relation
$\cF - \cF' - \cF''$ whenever these sheaves fit into an exact sequence. As coherent sheaves on well-behaved schemes allow for a finite locally free resolution, the property (\ref{multiplicativity}) allows for the extension of the definition of characteristic classes to all of $K(X)$.

For a complete scheme $X$ of dimension $n$, there is a degree map $A_n(X) \rightarrow \IZ$, given by $\deg(\sum n_i [Y_i]) = \sum n_i$, for $[Y_i]$ integral. The degree of an irreducible 0 dimensional subscheme $Y$ in $A_n(X)$ is hence its multiplicity, as defined above.

\vspace{0.5cm}
With these definitions, we can derive the Euler characteristic (\ref{chioy2}) of the structure sheaf of a subscheme $Y$ as follows. The ideal sheaf $\cI_Y$ associated to $Y$ is defined via the exact sequence
\be
\xymatrix{
0 ~\ar[r] & ~ \cI_Y ~ \ar[r] & ~\cO_X ~ \ar[r] & ~ \cO_Y ~ \ar[r] & ~ 0 \,.
}
\ee
By additivity of the Euler characteristic, one has
\ben
\chi(\cO_Y) = \chi(\cO_X) - \chi(\cI_Y) \,. \label{chioy}
\een
If we consider $\cI_Y$ as an abstract sheaf, forgetting about its embedding into $\cO_X$, then it is isomorphic to an invertible sheaf on $X$, i.e. a line bundle. This is simply the familiar correspondence between divisors $D=[Y]$ and line bundles $\cO_X(D)$,
\be
\cO_X(-D) = \cI_Y \,.
\ee
To calculate $\chi(\cO_Y)$ using eq.~(\ref{chioy}), we invoke the
Hirzebruch-Riemann-Roch theorem to compute $\chi(\cO_X)$ and
$\chi(\cO_X(-D))$. It states that {\em the Euler characteristic of  a
  locally free sheaf $\cE$ of rank $r$ on a nonsingular projective
  variety $X$ of dimension $n$ is given by $$\chi(\cE) = \deg
  \big(\ch(\cE)\cdot\td(X)\big)_n \,,$$ where $(\,-\,)_n$ denotes the
  component of degree $n$ in the Chow ring $A_*(X) \otimes \IQ$ and a dot denotes
  intersection product.}

When $X$ is a surface, the Todd class is given by
\ba
\td(X) = 1 - \mbox{$\frac{1}{2}$}\, K_X +
\mbox{$\frac{1}{12}$}\,\big(K_X^2 +c_2(X) \big) \ ,
\ea
where $K_X= - c_1(X)$ is the canonical divisor. Since
$\ch(\cO_X)=1$ one finds
\be
\chi(\cO_X)= \mbox{$\frac{1}{12}$}\,\big(K_X^2 +c_2(X) \big) \,.
\ee
For a line bundle $\cO_X(-D)$, one has $\ch(\cO_X(-D)) = \exp(-D)$. Therefore, $$\chi\big(\cO_X(-D) \big) = \mbox{$\frac{1}{2}$}\, D\cdot (D+K_X) +
\mbox{$\frac{1}{12}$}\, \big(K_X^2 + c_2(X) \big)\,.$$
Collecting these results, we arrive at formula (\ref{chioy2}).

\bigskip
Finally, we present a simple application of the definition of multiplicity presented above to the case of zero dimensional subschemes.

\vspace{0.3cm}

\begin{length_0_dim}  \label{length_0_dim}
The multiplicity of an irreducible zero dimensional subscheme $Y$ of a scheme $X$ of finite type over an algebraically closed field $k$ is given by $\dim_k(H^0(Y,\cO_Y))$. 
\end{length_0_dim}
\begin{proof}
The question is local, so we can assume $X$ affine with coordinate
ring $A$, and $Y= \spec A/I$. The generic point $y$ of $Y_{red}$ is $\sqrt{I}$ (this is prime by the irreducibility assumption). The local ring at the generic point is $\cO_{y,Y}= (A/I)_{\sqrt{I}}=A/I$. The second equality follows from $\dim A/I =0$. As the length of $A/I$ as a module over $k$ is equal to its dimension as a vector space, the lemma follows.
\end{proof}

\section{The Hilbert scheme of $\IC \IP^2$} \label{cp2}

The Hilbert scheme of hypersurfaces in projective space (zero sets of homogeneous polynomials in $\IP^r$) is easily obtained. For a hypersurface $Y$ of degree $d$, whose Hilbert polynomial is given by
\be
P^{\IP^r}_{Y}(t) = \binom{r+t}{r} - \binom{r+t-d}{r} \,,
\ee
it is given by projective space $\IP^N$, with $N= h^0(\IP^r, \cO_{\IP^r}(-d\,H))-1$ and $H$ the hyperplane divisor of $\IP^r$. Consider in particular the case $r=2$. As defined above, the hypersurfaces can be non-reduced and reducible, but they cannot include embedded points (as principal ideals do not possess embedded components).
Including such points increases the Euler characteristic of the subscheme. Hence the constant term of $P^{\IP^2}_{Y}(t)$, 
\be
P^{\IP^2}_{Y}(0) = \frac{3d-d^2}{2} \,,
\ee
is a lower bound for the Euler characteristic of a degree $d$ subscheme of $\IP^2$. In fact, this follows from a corollary of Hartshorne~\cite{HartshorneC}:

\vspace{0.5cm}

\begin{Hart}
\label{Hart}
Let $k$ be a field, $r>0$ an integer and $p\in \IQ[z]$ a numerical polynomial. Then a necessary and sufficient condition that $p$ be the Hilbert polynomial of a proper closed subscheme of $\IP^r_k$ is that when $p$ is written in the form
\be
p(z) = \sum_{t=0}^\infty \, \left[ \binom{z+t}{t+1} - \binom{z+t-m_t}{t+1} \right] \,,
\ee
one has $m_0 \ge m_1 \ge \ldots \ge m_{r-1} \ge 0$ and $m_r = m_{r+1} = \ldots = 0$.
\end{Hart}
In the case of interest, 
\be 
p(z) = m_1 \, z + \frac{2m_0 + m_1 - m_1^2}{2} \,,
\ee
from which our claim follows. 

This observation already suggests the factorization of the Hilbert scheme of curves on $\IP^2$ as
\be
\Hilb_{d\,t+n}^{\IP^2} = \Hilb_{d\,t+n_d}^{\IP^2} \times \big(\IP^2\big)^{[n-n_d]} \,,
\ee 
for $n \ge n_d = \frac{3d-d^2}{2}$. It follows from Corollary~\ref{Hart} that the left-hand side is empty for $n< n_d$.

At the level of the underlying topological spaces, this decomposition follows easily. With $S = \IC[x_0,x_1, x_2]$ the homogeneous coordinate ring of $\IP^2$, the subschemes of $\IP^2$ are in one-to-one correspondence with homogeneous ideals $I$ of $S$, via the map $I \mapsto \Proj S/I$. Since $S$ is a unique factorization domain, we can decompose $I$ uniquely into irreducible ideals
\be
I = \Big(\,\prod_i\, I^i_1\,\Big)~\Big(\, \prod_j\, I^j_0\,\Big) \,,
\ee
where the $I^i_1$ are generated by one element and correspond to subschemes of dimension~1, and the $I^j_0$ are generated by more than one element and correspond to subschemes of dimension~0.

\bibliography{biblio}
\bibliographystyle{utcaps}

\end{document}